\documentclass[twocolumn,epjc3]{svjour3}
\usepackage{color}
\usepackage{amssymb}
\usepackage{amsmath}
\usepackage{graphicx}
\usepackage{subcaption}
\journalname{Eur. Phys. J. C}
\usepackage{hyperref}
\hypersetup{
	colorlinks=true,        
	linkcolor=blue,         
	citecolor=cyan,         
}
\usepackage{graphicx}

\begin{document}

\title{Gravitational lensing for a boosted Kerr black hole\\ in the presence of plasma}

\author{
	Carlos A. Benavides-Gallego\thanksref{a},
	A.~A.~Abdujabbarov\thanksref{a,b} , 
and Cosimo~Bambi\thanksref{a,c,1}}

\institute{Center for Field Theory and Particle Physics and Department of Physics, Fudan University, 200438 Shanghai, China\label{a}
\and
Ulugh Beg Astronomical Institute,
	Astronomicheskaya 33, Tashkent  100052, Uzbekistan\label{b}
\and
Theoretical Astrophysics, Eberhard-Karls Universit\"at T\"ubingen, 72076 T\"ubingen, Germany\label{c}}


\thankstext{1}{\emph{e-mail:} bambi@fudan.edu.cn}

\maketitle

\begin{abstract}
We obtain the deflection angle for a boosted Kerr black hole in the weak field approximation. We also study the behavior of light in the presence of plasma by considering different distributions: singular isothermal sphere, non-singular isothermal gas sphere, and plasma in a galaxy cluster. We find that the dragging of the inertial system along with the boosted parameter $\Lambda$ affect the value of the deflection angle. As an application, we studied the magnification for both uniform and \textbf{SIS} distributions.     
\end{abstract}

\keywords{boosted Kerr metric, gravitational lensing, inhomogeneous plasma distributions}

\section{Introduction}\label{s:intro}

The revolutionary detection of gravitational waves from the coalescence of two black holes showed the formation of rapidly rotating black hole boosted with linear velocity~\cite{LIGO16b,LIGO16c,LIGO16d}. The possible observation of the electromagnetic counterpart from black hole merger could provide more information about angular and linear momentum of the black hole in such systems~\cite{Morozova14,Lyutikov11}. This fact indicates the importance of the inclusion of the boost parameter to Kerr spacetimes in order to study the effects of the boost velocity to the geometry (gravitational field) around a black hole. The solution of Einstein's vacuum field equations describing a boosted Kerr black hole relative to an asymptotic Lorentz frame at future null infinity was obtained in~\cite{Soares17}. The electromagnetic structure around a boosted black hole has been studied in~\cite{Morozova14}. The author of Ref.~\cite{Lyutikov11} has considered the solution of Maxwell equations in the background geometry of a boosted black hole. In the present paper, we study weak gravitational lensing around a boosted black hole described by the solution in~\cite{Soares17}. 

The gravitational lensing effect is a good tool to test Einstein's theory of general relativity. For a review on light propagation in the curved spacetime and geometrical optics in general relativity, see e.g.~\cite{Synge60,Schneider92,Perlick00,Perlick04}. The photon motion is also affected by the presence of a plasma and the effect of plasma around a compact objects on lensing effects has been studied in~\cite{Rogers15,Rogers17,Er18,Rogers17a,Broderick03,Bicak75,Kichenassamy85,Perlick17,Perlick15,Abdujabbarov17,Eiroa12,Kogan10,Tsupko10,Tsupko12,Morozova13,Tsupko14,Kogan17,Hakimov16,Turimov18,Benavides16,Kraniotis14}. In the literature, we can find a lot of work devoted to another optical property of black holes, the so-called black hole shadow~\cite{Vries00,Abdujabbarov17c,Abdujabbarov16a,Abdujabbarov16b,Abdujabbarov15,Abdujabbarov17b,Abdujabbarov15,Atamurotov15a,Amarilla12,Bambi09,Bambi12,Bambi13c,Ghasemi-Nodehi15,Cunha15,Wei13,Takahashi05,Ohgami15,Nedkova13,Falcke00,Grenzebach15,Takahashi04,Li14a,Wei15b,Papnoi14,Bambi10,Atamurotov13b,Atamurotov13,Tsukamoto14,Grenzebach2014,Abdujabbarov13c,Amarilla10,Amarilla13,Hioki09,Mizuno18}.

Starting from~\cite{Paczynski86a,Alcock93,Aubourg93,Udalski93,Paczynski96}, there is a rich literature on weak gravitational lensing. Strong gravitational lensing around spherically symmetric compact objects is described in~\cite{Tsupko09,Bozza06}.

In the present paper, we study weak lensing around a boosted black hole in the presence of plasma. The paper  is organized as follow. In Section~\ref{sec:Optics}, we briefly review the optics in curved spacetime and describe the procedure to obtain the deflection angle in the weak field approximation following~\cite{Kogan10,Morozova13}. In Section~\ref{sec:Kerr}, we present the boosted Kerr metric in both diagonal and non-diagonal cases (non-rotating and slowly rotating cases, respectively). In Subsections~\ref{sec:nonrotating_case_uniformplasma} and \ref{sec:Deflection_angle_for_the_slowly_rotating_case}, we find the expression for the deflection angle. Then, in section \ref{sec:Models}, we study the deflection angle in the presence of plasma, both for uniform and non-uniform distributions. For the inhomogeneous case, we consider three distribution models: singular isothermal sphere (\textbf{SIS}), non-singular isothermal sphere (\textbf{NSIS}), and the case of a plasma in a galaxy cluster (\textbf{PGC}). Finally, as an application, we devote section \ref{sec:magnification} to study the magnification for the uniform and \textbf{SIS} plasma distributions.\\

Throughout the paper we use the convention in which Greek indices run from 0 to 3, while Latin indices run from 1 to 3. Moreover, with the exception of Section~\ref{sec:Optics}, we use geometrized units, where $c=G=1$.

\section{\label{sec:Optics} Optics in a curved space-time}

In this section, we review the optics in a curved space-time developed by Synge in 1960~\cite{Synge60}. Let us consider a static spacetime metric describing a weak gravitational field in an asymptotically flat spacetime. The metric coefficients can be written as~\cite{Kogan10,Morozova13,Landau-Lifshitz2} 
\begin{eqnarray}
\label{asymptotically_flat}
g_{\alpha\beta}&=&\eta_{\alpha\beta}+h_{\alpha\beta}\ ,\\
g^{\alpha\beta}&=&\eta^{\alpha\beta}-h^{\alpha\beta}\ ,
\end{eqnarray}
where $\eta_{\alpha\beta}$ is the metric of the Minkowski spacetime, $h_{\alpha\beta}\ll 1$, $h_{\alpha\beta}\rightarrow 0$ for $x^\alpha \rightarrow \infty$, and $h^{\alpha\beta}=h_{\alpha\beta}$.\\

Using this approach for the static case, the phase velocity\footnote{The phase velocity is defined as the minimum value of~\cite{Synge60}
\begin{equation}
u'^2=1+\frac{dx_\alpha dx^\alpha}{(V_\beta dx^\beta)^2},
\end{equation}
where $u'$ is the velocity of a fictitious particle riding on the wave front relative to a time-like world-line $C$ (intersecting the wave) of an observer with 4-velocity $V^\mu$ (see \cite{Synge60} for details).} $u$ and the $4-$vector of the photon momentum $p^\alpha$ are related by the following equation~\cite{Synge60} 
\begin{equation}
\label{refraction_index_of_the_medium}
\frac{c^2}{u^2}=n^2=1+\frac{p_\alpha p^\alpha}{(p^0\sqrt{-g_{00}})^2}.\ 
\end{equation}
\\

In order to obtain the photon trajectories in the presence of a gravitational field, one can modify the Fermat's least action principle for the light propagation by considering a dispersive medium~\cite{Synge60}. Then, using the Hamiltonian formalism, it is easy to show that the variational principle%
\begin{equation}
\label{variational_principle}
\delta\left(\int p_\alpha dx^\alpha \right)=0\ ,
\end{equation}   
with the condition 
\begin{eqnarray}
\label{W}
W(x^\alpha,p_\alpha)=\frac{1}{2}\left[g^{\alpha\beta}p_\alpha p_\beta-(n^2-1)\left(p_0\sqrt{-g^{00}}\right)^2\right]=0,\nonumber
\end{eqnarray}
leads to the following system of differential equations that describes the trajectories of photons
\begin{eqnarray}
\label{differential_equations}
\frac{dx^\alpha}{d\lambda}&=&\frac{\partial W}{\partial p_\alpha}\ ,\nonumber\\
\frac{dp_\alpha}{d\lambda}&=&-\frac{\partial W}{\partial x^\alpha}\ ,
\end{eqnarray}
with the affine parameter $\lambda$ changing along the light trajectory. Note that the scalar function $W(x^\alpha,p_\alpha)$ has been defined by means of Eq.~(\ref{refraction_index_of_the_medium}).\\

In the Refs.~\cite{Kogan10,Morozova13}, it has been considered a static inhomogeneous plasma with a refraction index $n$ which depends on the space location $x^i$
\begin{eqnarray}
\label{refraction_index_inhomogeneous_plasma}
n^2&=&1-\frac{\omega_e^2}{[\omega(x^i)]^2}\ ,\\ \omega^2_e&=&\frac{4\pi e^2 N(x^i)}{m}=K_eN(x^i)\ , 
\end{eqnarray}
where $\omega(x^i)$ is the frequency of the photon that, due to gravitational redshift, depends on the space coordinates $x^1$, $x^2$, $x^3$, $e$ is the electron charge, $m$ is the electron mass, $\omega_e$ is the electron plasma frequency, and $N(x^i)$ is the electron concentration in an inhomogeneous plasma~\cite{Kogan10}.\\

According to Synge~\cite{Synge60}, for the case of a static medium in a static gravitational field, one can express the photon energy as 
\begin{equation}
p_0\sqrt{-g^{00}}=-\frac{1}{c}\hbar\omega(x^i).
\end{equation} 
Using Eq.~(\ref{refraction_index_inhomogeneous_plasma}) one can express the scalar function $W(x^\alpha,p_\alpha)$ in the following form
\begin{eqnarray}
\label{function_W}
W(x^\alpha,p_\alpha)=\frac{1}{2}\left[g^{\alpha\beta}p_{\alpha}p_{\beta}+\frac{\omega^2_e\hbar^2}{c^2}\right],
\end{eqnarray} 
where $\hbar$ is the Planck's constant. The scalar function expressed in Eq.~(\ref{function_W}) has been used in Refs.~\cite{Kogan10,Morozova13} to find the equations of light propagation for diagonal and non-diagonal spacetimes.\\
 
In contrast with the case of a flat spacetime in vacuum, where the solution for photon's trajectory is a straight line, the presence of an arbitrary medium in curved spacetimes makes photons move along bent trajectories. However, taking into account only small deviations, it is possible to use the components of the 4-momentum of the photon moving in a straight line along the $z-$axis as an approximation. This components are given by (see, e.g.~\cite{Kogan10,Morozova13})  
\begin{eqnarray}
\label{null_approximation}
p^\alpha&=&\left(\frac{\hbar\omega}{c},0,0,\frac{n\hbar\omega}{c}\right)\ ,\\p_\alpha&=&\left(-\frac{\hbar\omega}{c},0,0,\frac{n\hbar\omega}{c}\right).\label{null_approximation2}
\end{eqnarray}
Eqs.~(\ref{null_approximation}) and (\ref{null_approximation2}) are known as the null approximation. It is important to point out that both $\omega$ and $n$ are evaluated at $\infty$. In this sense, we have introduced the notation in which
\begin{equation}
\begin{aligned}
\omega&=\omega(\infty)\\
n&=n(\infty).\\
\end{aligned}
\end{equation}             
This notation has been also used in \cite{Kogan10,Morozova13}, and will be used along the manuscript.

\subsection{\label{sec:level3}Equations of light propagation in a diagonal spacetime}

First, we consider the spacetime with a diagonal metric tensor.  In this spacetime, the components of the metric tensor $g_{\alpha\beta}$ vanish for $\alpha\ne \beta$. Hence, after using Eq.~(\ref{function_W}), the system in~(\ref{differential_equations}) can be expressed as \cite{Kogan10}

\begin{equation}
\label{systme_diagonal}
\begin{aligned}
\frac{dx^i}{d\lambda}&=g^{ij}p_j,\\
\frac{dp_i}{d\lambda}&=-\frac{1}{2}g^{lm}_{\;\;\;\;,i}p_l p_m-\frac{1}{2}g^{00}_{\;\;\;\;,i}p^2_0-\frac{1}{2}\frac{\hbar^2}{c^2}K_eN_{,i}.
\end{aligned}
\end{equation}

Then, with the aid of the null approximation, the first equation in~(\ref{systme_diagonal}) reduces to
\begin{equation}
\label{relation_differential}
\frac{dz}{d\lambda}=\frac{n\hbar\omega}{c}\ .
\end{equation}  
In the null approximation, the $3-$vector in the direction of the photon momentum is written as $e^i =e_i=(0,0,1)$. Therefore $p_i$ can be expressed as
\begin{equation}
p_i=\frac{n\hbar\omega}{c}(0,0,1)=\frac{n\hbar\omega}{c}e_i.
\end{equation}
Hence, the second equation in (\ref{systme_diagonal}) can be expressed by 
\begin{eqnarray}
\frac{d}{d\lambda}\left(\frac{n\hbar\omega}{c}e_i\right)&=&-\frac{1}{2}g^{lm}_{\;\;\;\;,i}p_l p_m\nonumber\\&&-\frac{1}{2}g^{00}_{\;\;\;\;,i}p^2_0-\frac{1}{2}\frac{\hbar^2}{c^2}K_eN_{,i}.
\end{eqnarray}
Then, after using Eq.~(\ref{relation_differential}) and differentiating, the last expression takes the form 
\begin{eqnarray}
\label{second_equation_system}
\frac{de_i}{dz}&=&-\frac{1}{2}\frac{c^2}{n\hbar^2\omega^2}\left(g^{00}_{\;\;\;\;,i}(p_0)^2+g^{lm}_{\;\;\;\;,i}p_l p_m+\frac{\hbar^2}{c^2}K_eN_{,i}\right)\nonumber\\
&&-e_i\frac{dn}{dz}\ .
\end{eqnarray}
In Ref.\cite{Kogan10}, only those components of the $3-$vector that are perpendicular to the initial direction of propagation were taken into account. In this sense, the contribution to the deflection of photons is due only to the change in $e_1$ and $e_2$. Hence, after using the null approximation $e_i=0$ along with the assumption of weak gravitational field, Eq.~(\ref{second_equation_system}) reduces to  
\begin{equation}
\label{derivative_e}
\frac{de_i}{dz}=\frac{1}{2}\left(h_{33,i}+\frac{1}{n^2}h_{00,i}-\frac{1}{n^2\omega^2}K_eN_{,i}\right)\ ,
\end{equation}   
for $i=1,2$.\\

The deflection angle is determined by the change of the $3-$vector $e_i$. This means that 
\begin{equation}
\vec{\hat{\alpha}}=\mathbf{e}(+\infty)-\mathbf{e}(-\infty).
\end{equation} 
Then, using Eq.~(\ref{derivative_e}), the deflection angle becomes
\begin{eqnarray}
\hat{\alpha}_i=\frac{1}{2}\int^{\infty}_{-\infty}\left(h_{33,i}+\frac{\omega^2}{\omega^2-\omega^2_e}h_{00,i}-\frac{K_e}{\omega^2-\omega^2_e}N_{,i}\right)dz,\nonumber\\\label{deflection_angle_diagonal_weak_field}
\end{eqnarray}
for $i=1,2$. In the last expression $\omega_e$ and $n$ are evaluated at infinity, and $\omega(\infty)=\omega$ \cite{Kogan10}. In terms of the impact parameter $b$, Eq.~(\ref{deflection_angle_diagonal_weak_field}) takes the form \cite{Kogan10}
\begin{eqnarray}
\label{deflection_angle_diagonal_weak_field_b}
\hat{\alpha}_b&=&\frac{1}{2}\int^\infty_{-\infty}\frac{b}{r}\nonumber \\
&&\left(\frac{dh_{33}}{dr}+\frac{1}{1-\omega^2_e/\omega^2}\frac{dh_{00}}{dr}-\frac{K_e}{\omega^2-\omega^2_e}\frac{dN}{dr}\right)\ ,
\end{eqnarray}
where $r=\sqrt{b^2+z^2}$.

\subsection{Equations of light propagation in a non-diagonal spacetime}

Now we consider a spacetime with a non-diagonal metric tensor; that is, the components of metric tensor  $g_{\alpha\beta}$ do not vanish for $\alpha\neq\beta$. Therefore, the scalar function $W(x^\alpha,p_\alpha)$ in Eq.~(\ref{function_W}) can be expressed as \cite{Morozova13}
\begin{eqnarray}
\label{function_W_for_non_diagonal_space_time}
W&(&x^\alpha,p_\alpha)=\nonumber\\
&&\frac{1}{2}\left[g^{00}p^2_0+2g^{0l}p_{0}p_{l}+g^{lm}p_{l}p_{m}+\frac{\omega^2_e\hbar^2}{c^2}\right].
\end{eqnarray}  
Hence, the system of differential equations in (\ref{differential_equations}) takes the form
\begin{eqnarray}
\frac{dx^i}{d\lambda}&=&g^{ij}p_j\\
\frac{dp_i}{d\lambda}&=&-\frac{1}{2}g^{lm}_{\;\;\;\;,i}p_l p_m-\frac{1}{2}g^{00}_{\;\;\;\;,i}p^2_0-g^{0l}_{\;\;\;\;,i}p_0p_l\nonumber\\
&&-\frac{1}{2}\frac{\hbar^2}{c^2}K_eN_{,i}.
\end{eqnarray}
Then, using Eq.~(\ref{relation_differential}) and assuming that the gravitational field is weak, we obtain 
\begin{eqnarray}
\frac{dp_i}{dz}&=&\frac{1}{2}\frac{n\hbar\omega}{c}\nonumber \\
&&\times\left(h_{33,i}+\frac{1}{n^2}h_{00,i}+\frac{1}{n}h_{03,i}-\frac{K_eN_{,i}}{n^2\omega^2}\right).
\end{eqnarray}
Therefore, following the procedure in Subsection~\ref{sec:level3}, the deflection angle for a non-diagonal spacetime in the weak limit has the form
\begin{eqnarray}
\label{deflection_angle_non_diagonal}
\hat{\alpha}_i&=&\frac{1}{2}\int^{\infty}_{-\infty}\bigg(h_{33,i}+\frac{\omega^2}{\omega^2-\omega^2_e}h_{00,i}+\frac{1}{n}h_{03,i}\nonumber\\
&&-\frac{K_eN_{,i}}{\omega^2-\omega^2_e}\bigg)dz\ .
\end{eqnarray}


\section{\label{sec:Kerr} Boosted Kerr metric}
The boosted Kerr metric, which describes a boosted black hole relative to an asymptotic Lorentz frame, is a solution of Einstein's vacuum field equations obtained in~\cite{Soares17}. This solution has three parameters: mass, rotation and boost. In Kerr-Schild coordinates, the line element reads
\begin{eqnarray}
\label{boosted_kerr_metric_in_kerrSchild_coordinates}
ds^2&=&-\left(1-\frac{2Mr}{\Sigma}\right)dt'^2+\left(1+\frac{2Mr}{\Sigma}\right)dr^2\nonumber+\frac{\Sigma}{\Lambda}d\theta^2\\
&&+\frac{A\sin^2(\theta)}{\Lambda^2\Sigma}d\phi^2-\frac{4Mra\sin^2\theta}{\Lambda\Sigma}dt'd\phi\nonumber\\
&&-\frac{4Mr}{\Sigma}dt'dr-\frac{2a\sin^2\theta}{\Lambda}\left(1-\frac{2Mr}{\Sigma}\right)drd\phi\ ,
\end{eqnarray}  
with
\begin{eqnarray}
\label{definitions}
\Sigma&=&r^2+a^2\left(\frac{\beta+\alpha\cos\theta}{\alpha+\beta\cos\theta}\right)^2\ ,\\
\Lambda&=&(\alpha+\beta\cos\theta)^2\ ,\\
A&=&\Sigma^2+a^2\left(\Sigma+2Mr\right)\sin^2\theta\ ,
\end{eqnarray}
where $a={J}/{M}$ is the specific angular momentum of the compact object with total mass $M$, $\alpha=\cosh\gamma$, $\beta=\sinh\gamma$, and $\gamma$ is the usual Lorentz factor which defines the boost velocity $v$ by the formula $v=\tanh\gamma={\beta}/{\alpha}$. Note that the metric in~(\ref{boosted_kerr_metric_in_kerrSchild_coordinates})
exactly reduces to the Kerr one when $\Lambda=1$ ($v=0$). It is also important to point out that the direction of the boost for the Kerr black hole is along the axis of rotation while for Schwarzschild is along the $z-$axis.\\
 
To study the deflection angle for the boosted Kerr metric in the presence of a medium, we consider both the non-rotating and the slowly rotating cases. In this sense, following the ideas in~\cite{Kogan10} and \cite{Morozova13}, we devote this section to find the form of the line element (\ref{boosted_kerr_metric_in_kerrSchild_coordinates}) in each case.

\subsection{\label{sec:kerr_nonrotating}Boosted Kerr metric: non-rotating case}
The non-rotating case is obtained by setting $a=0$. Hence, the metric (\ref{boosted_kerr_metric_in_kerrSchild_coordinates}) reduces to 
\begin{eqnarray}
\label{boosted_kerr_metric_in_kerrSchild_coordinates_a0}
ds^2&=&-\left(1-\frac{2M}{r}\right)dt'^2+\left(1+\frac{2M}{r}\right)dr^2\nonumber\\
&&+\frac{r^2}{\Lambda}d\theta^2+\frac{\sin^2\theta}{\Lambda^2}r^2d\phi^2-\frac{4M}{r}dt'dr\ .
\end{eqnarray} 

In Ref.~\cite{Kogan10}, Cartesian coordinates have been used to find the terms $h_{ik}$. Nevertheless, before changing the coordinates, we want to write the form of the metric in Eq.~(\ref{boosted_kerr_metric_in_kerrSchild_coordinates_a0}) for small values of the velocity ($v\ll1$). In order to do so, we express $1/\Lambda$ and $1/\Lambda^2$ in terms of $v$ and consider a Taylor expansion up to first order. Therefore, the metric (\ref{boosted_kerr_metric_in_kerrSchild_coordinates_a0}) takes the form
\begin{eqnarray}
\label{final_line_element}
ds^2&=&-\left(1-\frac{2M}{r}\right)dt'^2+\left(1+\frac{2M}{r}\right)dr^2\nonumber\\
&&+r^2(1-2v\cos\theta)d\theta^2+r^2\sin^2\theta d\phi^2\nonumber\\
&&-4vr^2\sin^2\theta\cos\theta d\phi^2-\frac{4M}{r}dt'dr.
\end{eqnarray}
Now, to transform the line element (\ref{final_line_element}) into Boyer-Lindquist coordinates, we use the relation (see \cite{Visser07}, page 15)
\begin{equation}
\label{Boyer_Lindquist}
t'=t-2M\ln\left(\frac{r}{2M}-1\right);
\end{equation}
from which one can easily obtain 
\begin{eqnarray}
\label{line_element_in_Boyer_lindquist_coordinate}
ds^2&=&-\left(1-\frac{2M}{r}\right)dt^2+\left(1-\frac{2M}{r}\right)^{-1}dr^2\nonumber\\
&&r^2\left[(1-2v\cos\theta)d\theta^2+(1-4v\cos\theta)\sin^2\theta d\phi^2\right].\nonumber\\
\end{eqnarray}

In the weak field limit, the approximation is done by considering ${2M}/{r}\ll 1$. In this sense, according to \cite{Kogan10}, the main idea is to express the line element in Eq.~(\ref{line_element_in_Boyer_lindquist_coordinate}) as
\begin{equation}
\label{weak_field_limit}
ds^2=ds^2_0+ds'^2
\end{equation}
where  
\begin{equation}
ds^2_0=-dt^2+dr^2+r^2(d\theta^2+\sin^2\theta d\phi^2),
\end{equation}
is the flat space-time, and $ds'^2$ is the part of the metric containing the perturbation terms $h_{ik}$. Therefore, after considering the weak approximation, the line element~(\ref{line_element_in_Boyer_lindquist_coordinate}) has the form
\begin{eqnarray}
\label{line_element_in_weak_limit_in_Boyer_lindquist_coordinates}
ds^2&=&ds^2_0+\frac{2M}{r}dt^2+\frac{2M}{r}dr^2 \nonumber\\
&&-2vr^2\cos\theta d\theta^2-4vr^2\cos\theta\sin^2\theta d\phi^2.
\end{eqnarray}
Eq.~(\ref{line_element_in_Boyer_lindquist_coordinate}) is the non-rotating boosted Kerr  metric in the weak field approximation expressed in Boyer-Lindquist coordinates. In order to identify the components $h_{ik}$, we need to express the line element in Eq.~(\ref{line_element_in_weak_limit_in_Boyer_lindquist_coordinates}) in Cartesian coordinates. After following the procedure described in Appendix~I, we found that $h_{00}$ and $h_{33}$ are
\begin{eqnarray}
\label{perturbation_h00_h33}
h_{00}&=&\frac{2M}{r}\\
\label{perturbation_h33}
h_{33}&=&\frac{2M}{r}\cos^2\theta-2v\cos\theta\sin^2\theta,
\end{eqnarray}

\subsection{\label{sec:kerr_rotating} Boosted Kerr metric: rotating case}

The spacetime describing a slowly rotating massive object was obtained in~\cite{Hartle68}. However, in this work, we use the form of the metric reported in~\cite{Morozova13}. Using geometrized units, this metric takes the form 
\begin{eqnarray}
ds^2&=&-\left(1-\frac{2M}{r}\right)dt^2+\left(1-\frac{2M}{r}\right)^{-1}dr^2\nonumber \\
&&+r^2(d\theta^2+\sin^2\theta d\phi^2)-2\omega_{LT}r^2\sin^2\theta dtd\phi\ ,
\end{eqnarray}
where $\omega_{LT}={2Ma}/{r^3}={2J}/{r^3}$ is the Lense-Thirring angular velocity of the dragging of inertial frames. \\

For the case of the boosted Kerr metric, the line element has the same form. Introducing the notation  $\overline{\omega}_{LT}={2\overline{J}}/{r^3}$, where $\overline{J}={J}/{\Lambda}$, one may obtain the ``modified'' metric of slowly rotating boosted velocity. Finally, the spacetime around  boosted slowly rotating objects can be expressed by the following metric%
\begin{eqnarray}
\label{boosted_rotating_case}
ds^2&=&-\left(1-\frac{2M}{r}\right)dt^2+\left(1-\frac{2M}{r}\right)^{-1}dr^2\nonumber\\
&&+r^2(d\theta^2+\sin^2\theta d\phi^2)-2\overline{\omega}_{LT}r^2\sin^2\theta dtd\phi.
\end{eqnarray} 
When $v=0$, the expression in~(\ref{boosted_rotating_case}) reduces to that in~\cite{Morozova13}.  

\section{Deflection angle in uniform plasma}

\subsection{\label{sec:nonrotating_case_uniformplasma}Deflection of light for the non-rotating case} 
In Subsection~\ref{sec:level3}, we discussed the procedure in~\cite{Kogan10} to obtain Eq.~(\ref{deflection_angle_diagonal_weak_field_b}). Now, we apply this result to find the deflection angle for the boosted Kerr metric in the presence of a uniform plasma. We first consider the non-rotating case. From Eqs.~(\ref{perturbation_h00_h33}) and (\ref{perturbation_h33}) we have that
\begin{eqnarray}
\label{h33_derivative_with_respect_to_r}
\frac{b}{r}\frac{dh_{00}}{dr}&=&-\frac{2Mb}{r^3}=-\frac{2Mb}{\sqrt{b^2+z^2}^\frac{3}{2}}\ ,\\
\frac{b}{r}\frac{dh_{33}}{dr}&=&-\frac{6Mb}{r^5}z^2+\frac{2bv}{r^3}z-\frac{6vb}{r^5}z^3\nonumber \\
&=&-\frac{6Mb}{(b^2+z^2)^\frac{5}{2}}z^2+\frac{2bv}{(b^2+z^2)^\frac{3}{2}}z\nonumber\\
&&-\frac{6vb}{(b^2+z^2)^\frac{5}{2}}z^3\ .
\end{eqnarray}
Then, recalling that $\cos\theta={z}/{r}$, $r=\sqrt{b^2+z^2}$, and using Eq.~(\ref{deflection_angle_diagonal_weak_field_b}), the deflection angle is 
\begin{eqnarray}
\label{delflection_angle_with_plasma_boosted_kerr_metric}
\hat{\alpha}_b&=&-3Mb\int^\infty_{-\infty}\frac{z^2}{(b^2+z^2)^\frac{5}{2}}dz\nonumber\\
&&+bv\int^\infty_{-\infty}\frac{z}{(b^2+z^2)^\frac{3}{2}}dz\nonumber\\
&&-3bv\int^\infty_{-\infty}\frac{z^3}{(b^2+z^2)^\frac{5}{2}}dz\nonumber\\
&&-Mb\int^\infty_{-\infty}\frac{\omega^2}{(\omega^2-\omega^2_e)(b^2+z^2)^\frac{3}{2}}dz\nonumber\\
&&-\frac{bK_e}{2}\int^\infty_{-\infty}\frac{1}{\omega^2-\omega^2_e}\frac{1}{r}\frac{dN}{dr}dz\ .
\end{eqnarray}
Thus, after integration, we obtain 
\begin{eqnarray}
\label{deflection_angle_nonrotating_case}
\hat{\alpha}_b&=&\frac{2M}{b}+\frac{2Mb}{1-\frac{\omega^2_e}{\omega^2}}\int^\infty_{0}\frac{dz}{(b^2+z^2)^\frac{3}{2}}\nonumber\\
&&+\frac{bK_e}{2}\int^\infty_{-\infty}\frac{1}{\omega^2-\omega^2_e}\frac{1}{r}\frac{dN}{dr}dz.
\end{eqnarray}
In the last expression, we took into account the symmetry of the limits (see appendix II for details). We also considered the fact that the deflection angle is defined as the difference between the initial and the final ray directions; that is, $\mathbf{\hat{\alpha}}=\mathbf{e}_{in}-\mathbf{e}_{out}$. Therefore, it has the opposite sign (see \cite{Schneider92}).\\

From Eq.~(\ref{deflection_angle_nonrotating_case}) we note that, at first order, $\hat{\alpha}_b$ does not depend on the velocity. This is due to the fact that the second and third integrals in Eq.~(\ref{delflection_angle_with_plasma_boosted_kerr_metric}), which contain the dependence on $v$, vanish. If we consider a uniform plasma ($\omega_e$ constant), and the approximation $1-n\ll\frac{\omega_e}{\omega}$, Eq.~(\ref{deflection_angle_nonrotating_case}) reduces to \cite{Kogan10}\\

\begin{eqnarray}
\label{non_rotating_case_approximation}
\hat{\alpha}_b=\frac{2M}{b}\left(1+\frac{1}{1-\frac{\omega^2_e}{\omega^2}}\right).
\end{eqnarray} 

\begin{figure*}[t]
\includegraphics[scale=0.38]{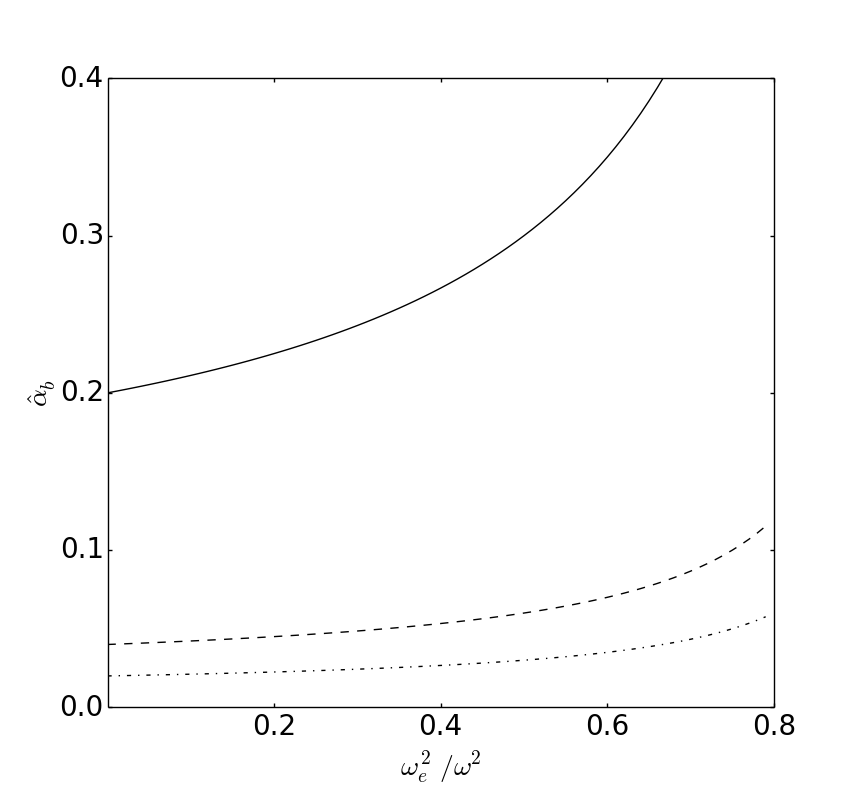}
\includegraphics[scale=0.38]{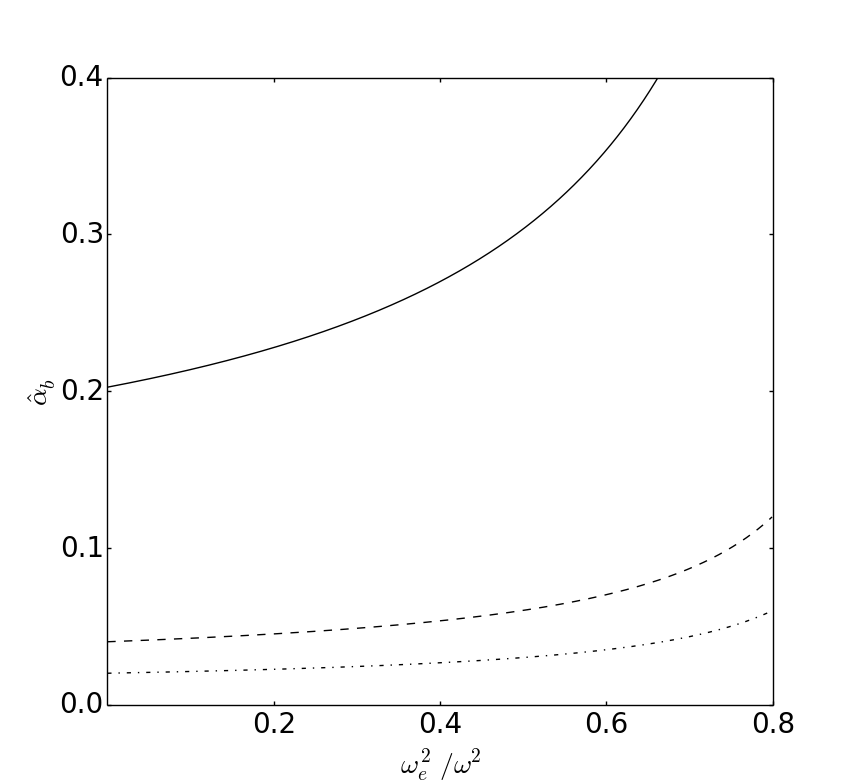}
\caption{\textbf{left}: Plot of $\hat{\alpha}_b$ vs. $\omega^2_e/\omega^2$ for $b/2M=10$ (continuous line), $b/2M=50$ (dashed line), and $b/2M=100$ (dot-dashed line) for uniform plasma. \textbf{right}: Plot of $\hat{\alpha}_b$ vs. $\omega^2_e/\omega^2$ for the rotating case. We used different values of the impact parameter: $b/2M=10$ (continuous line), $b/2M=50$ (dashed line), and $b/2M=100$ (dot-dashed line).  We assumed $\Lambda=0.5$, $J_r/M^2=0.25$, $\sin\chi=1$, and $\omega^2_e/\omega^2=0.5$. Note that there is a small increment for $b/2M=10$ when we compare with  Schwarzschild (left panel). \label{fig2}}
\end{figure*}

In Fig.~\ref{fig2} left we plotted $\hat{\alpha}_b$ as a function of $\omega^2_e/\omega^2$ for different values of $b/2M$. The plot shows that $\hat{\alpha}_b$ increases as the ration $\omega^2_e/\omega^2$ increases. On the other hand, for small values of $b/2M$ the values of the deflection angle are greater. For example, for $b/2M=100$ the figure shows that $\hat{\alpha}_b$ is greater than $0.2$; however, for $b/2M=50,100$ the deflection angle is less than $0.1$. It is also possible to see from the figure that $\hat{\alpha}_b$ has the value $4M/b$ when there is not plasma ($\omega_e=0$).

\subsection{\label{sec:Deflection_angle_for_the_slowly_rotating_case} Deflection angle for the slowly rotating case}
\begin{figure}[h!]
\begin{centering}
\includegraphics[scale=0.38]{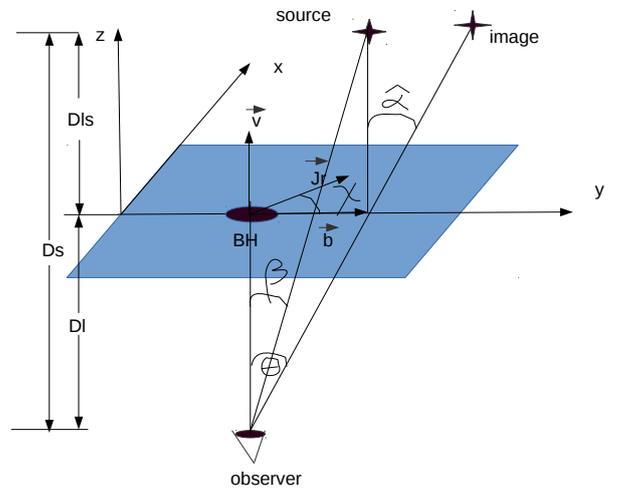}
\label{alpha_vs_b2M_uniform_plasma_rotating_case}
\caption{Schematic representation of the gravitational lensing system.  Here, $\chi$ represents the inclination angle between the vectors $\mathbf{J}_r$ and $\mathbf{b}$. In the figure $D_s$, $D_{l}$, and $D_{ls}$ are the distances from the source to the observer, from the lens to the observer, and from the source to the lens, respectively. \label{esquema}}
\end{centering}
\end{figure}

Due to the presence of non-diagonal terms in the line element (\ref{boosted_rotating_case}), we use the form of the deflection angle in Eq.~(\ref{deflection_angle_non_diagonal}). According to \cite{Morozova13}, the effect of dragging of the inertial frame contributes to $\hat{\alpha}$ only by means of the projection $\overline{J}_r$ of the angular momentum. Hence, after the introduction of polar coordinates $(b,\chi)$ on the intersection point between the light ray and the $xy$-plane, where $\chi$ is the angle between $\vec{J}_{r}$ and $\vec{b}$
, we find that~\cite{Morozova13}~(see Fig.~\ref{esquema})

\begin{equation}
\label{h03_rotating_case}
h_{03}=-2\frac{\overline{J}_rb\sin\chi}{(b^2+z^2)^{3/2}}.
\end{equation}
Since Eq.~(\ref{h03_rotating_case}) depends on $\chi$ and $b$, the deflection angle contains two contributions: the partial derivatives 
\begin{eqnarray}
\label{derivative_of_h03_b_and_chi}
\frac{\partial h_{03}}{\partial b}&=&-2\overline{J}_r\sin\chi\left(\frac{1}{(b^2+z^2)^{3/2}}-\frac{3b^2}{(b^2+z^2)^{5/2}}\right) ,\\
\frac{\partial h_{03}}{\partial\chi}&=&-2\frac{\overline{J}_rb\cos\chi}{(b^2+z^2)^{3/2}}\ .
\end{eqnarray}   
Thus, Eq.~(\ref{deflection_angle_non_diagonal}), for both contributions, takes the form
\begin{eqnarray}
\label{deflection_angle_rotating_case}
\hat{\alpha}_b&=&\hat{\alpha}_{bS}-2\overline{J}_r\sin\chi\nonumber \\
&&\times\int^\infty_{0}\left(\frac{1}{n(b^2+z^2)^{3/2}}-\frac{3b^2}{n(b^2+z^2)^\frac{5}{2}}\right)dz\\
\hat{\alpha}_\chi&=&-2\overline{J}_r\cos\chi\int^\infty_{0}\frac{1}{n(b^2+z^2)^{3/2}}dz ,
\end{eqnarray}
where $\hat{\alpha}_{bS}$ is the deflection angle for Schwarzschild (see Eq.~(\ref{deflection_angle_nonrotating_case})). Therefore, considering an homogeneous plasma (constant value of $\omega_e$), these contributions reduce to
\begin{eqnarray}
\label{deflection_angle_rotating_case_homgeneous_plasma}
\hat{\alpha}_b&=&\underbrace{\frac{2M}{b}\left(1+\frac{1}{1-\frac{\omega^2_e}{\omega^2}}\right)}_{\hat{\alpha}_{bS}}+\underbrace{\frac{1}{\sqrt{1-\frac{\omega^2_e}{\omega^2}}}\frac{2J_r\sin\chi}{b^2\Lambda}}_{\hat{\alpha}_{bD}}\\
\label{chi}
\hat{\alpha}_\chi&=&-\frac{2J_r\cos\chi}{b^2\Lambda\sqrt{1-\frac{\omega^2_e}{\omega^2}}};
\end{eqnarray}
where $n$ was replaced by $\sqrt{1-\frac{\omega^2_e}{\omega^2}}$. It is important to point out that Eq.~(\ref{deflection_angle_rotating_case_homgeneous_plasma}) is only valid for $\omega>\omega_e$, because waves with $\omega<\omega_e$ do not propagate in the plasma~\cite{Kogan10,Ginzburg70}. \\ 

\begin{figure}[ht]
\includegraphics[scale=0.38]{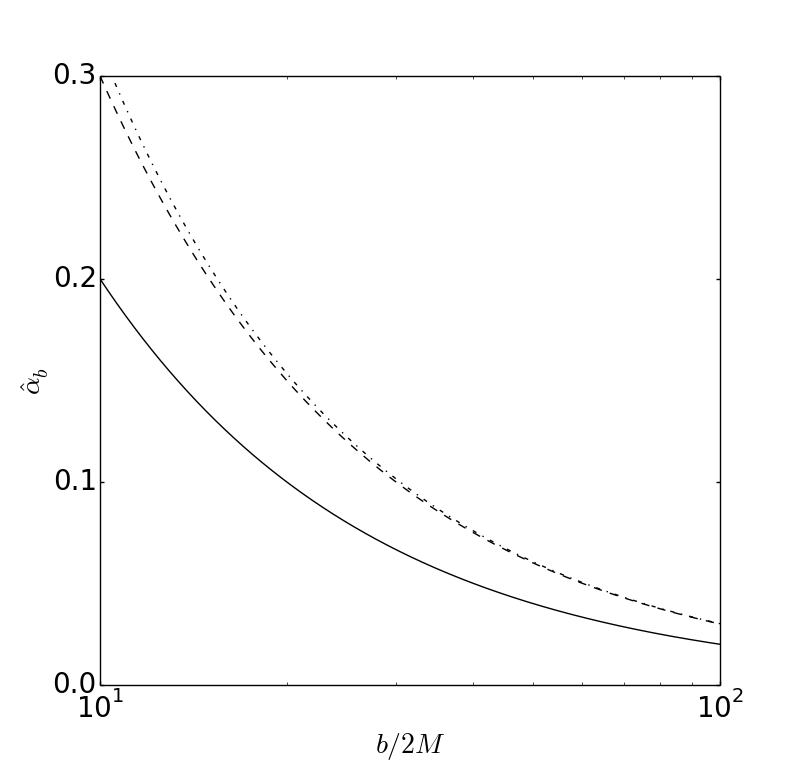}
\label{alpha_vs_b2M_uniform_plasma_rotating_case}
\caption{Plot of $\hat{\alpha}_b$ vs. $b/2M$ in the presence of uniform plasma for the slowly rotating (dot-dashed line) and $\hat{\alpha}_{bS}$ (dashed line). In the figure it is also plotted the Schwarzschild case in vacuum (continuous line). We used $\Lambda=0.5$, $J_r/M^2=0.25$, $\sin\chi=1$, and $\omega^2_e/\omega^2=0.5$. \label{fig3}}
\end{figure}


\begin{figure}[h!]
\includegraphics[scale=0.35]{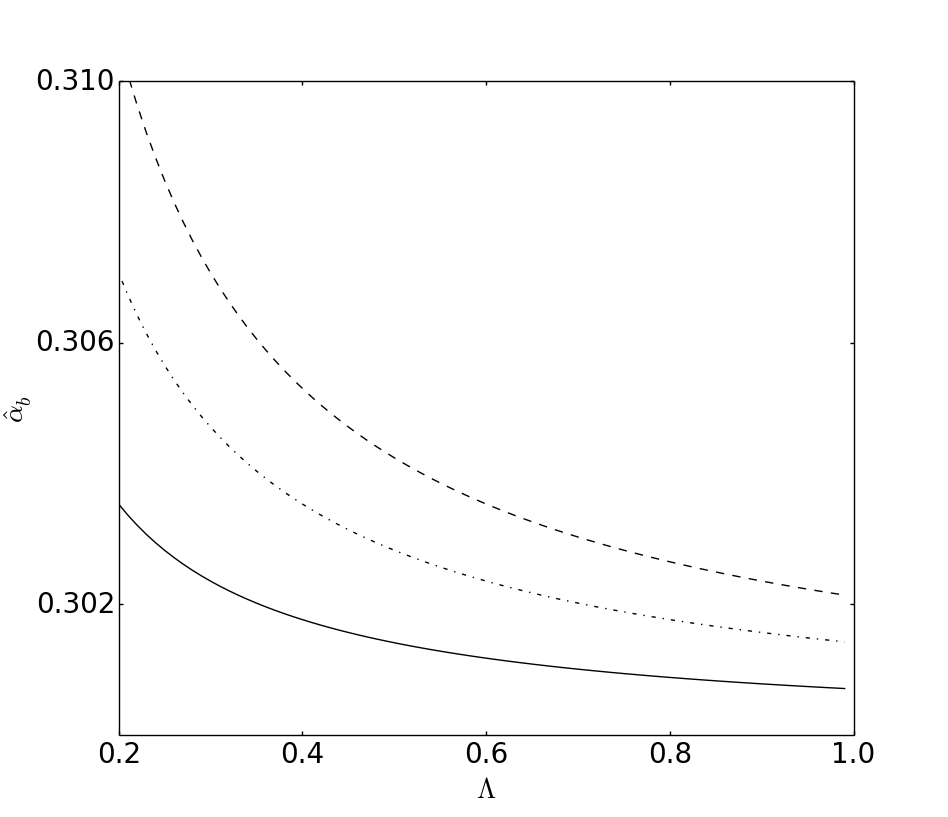}
\label{rotating_Lambda}
\caption{Plot of $\hat{\alpha}_b$ vs. $\Lambda$ for $J_r/M^2=0.1$ (continuous line), $J_r/M^2=0.2$ (dot-dashed line) , and $J_r/M^2=0.3$ (dashed line). We assumed  $b/2M=10$, $\sin\chi=1$, and $\omega^2_e/\omega^2=0.5$. \label{fig5}}  
\end{figure}

In Fig.~\ref{fig3}, we plot $\hat{\alpha}_{bS}$ and $\hat{\alpha}_b$ for the slowly rotating case as a function of the impact parameter $b/2M$. From this figure, we can see that there is a difference between both angles. This means that $\hat{\alpha}_b$ for a boosted Kerr black hole is greater than $\hat{\alpha}_{bS}$. This is due to the rotation and boost velocity $v$, which is larger for small values of $b/2M$. On the other hand, for larger values of the impact parameter $b/2M$, this difference becomes very small, and both angles behave in the same way since ${2J_r\sin\chi}/(nb^2\Lambda)\rightarrow 0$ when $b/2M\rightarrow\infty$.\\

Fig.~\ref{fig2} right shows $\hat{\alpha}_b$ as a function of $\omega^2_e/\omega^2$. The behavior is very similar to that of Schwarzschild (Fig.~\ref{fig2} left). However, note that there is a small increment for $b/2M=10$. On the other hand, we see that the deflection angle tends to  $2M/b+{2J_r\sin\chi}/(b^2\Lambda)$ when there is not plasma ($\omega^2_e=0$). 

In Fig.~\ref{fig5} we plotted Eq.~(\ref{deflection_angle_rotating_case_homgeneous_plasma}) as a function of $\Lambda$ for different values of $J_r$. We took into account the condition in which $0<\Lambda\leq 1$ in order to give the values. In this figure, for different values of $\Lambda$, we see that $\hat{\alpha}_b$ is bigger when $\Lambda\rightarrow 0$. Moreover, for $\Lambda=1$, the deflection angle reduces to the value $\hat{\alpha}_{bS}+{2J_r\sin\chi}/{nb^2}$.


\section{\label{sec:Models} Models for the boosted Kerr metric with non-uniform plasma distribution}

The deflection angle for a boosted Kerr metric in a non-uniform plasma was calculated in Subsection~\ref{sec:Deflection_angle_for_the_slowly_rotating_case}. Hence, for  ${\omega^2_e}/{\omega^2}\ll 1$, Eq.~(\ref{deflection_angle_rotating_case}) reduces to 
\begin{eqnarray}
\label{non_uniform_plasma}
\hat{\alpha}_b&=&\underbrace{\frac{4M}{b}}_{\hat{\alpha}_{S1}}+\underbrace{\frac{2Mb}{\omega^2}\int^\infty_0\frac{\omega^2_e}{r^3}dz}_{\hat{\alpha}_{S2}}\nonumber\\
&&+\underbrace{\frac{bK_e}{\omega^2}\int^\infty_0\frac{1}{r}\frac{dN}{dr}dz}_{\hat{\alpha}_{S3}}+\underbrace{\frac{bK_e}{\omega^4}\int^\infty_0\frac{\omega^2_e}{r}\frac{dN}{dr}dz}_{\hat{\alpha}_{S4}}\nonumber\\
&&+\underbrace{\frac{2J_r}{\Lambda b^2}\sin\chi}_{\hat{\alpha}_{B1}}-\underbrace{\frac{J_r}{\Lambda\omega^2}\sin\chi\int^\infty_0\frac{\omega_e^2}{r^3}dz}_{\hat{\alpha}_{B2}}\nonumber\\
&&+\underbrace{\frac{3b^2J_r}{\Lambda\omega^2}\sin\chi\int^\infty_0\frac{\omega_e^2}{r^5}dz}_{\hat{\alpha}_{B3}},
\end{eqnarray}
where $r=\sqrt{b^2+z^2}$, and $S$ and $B$ stand for Schwarzschild and Boosted, respectively. Using Eq.~(\ref{non_uniform_plasma}), we calculate the deflection angle by considering different plasma distributions: singular isothermal sphere (\textbf{SIS}), non-singular isothermal gas sphere (\textbf{NSIS}),  and a plasma in a galaxy cluster (\textbf{PGC}).\\

Eq.~(\ref{non_uniform_plasma}) is quite similar to that obtained in~\cite{Kogan10}. In this equation, we also find the vacuum gravitational deflection $\hat{\alpha}_{S1}$, the correction to the gravitational deflection due to the presence of the plasma $\hat{\alpha}_{S2}$, the refraction deflection due to the inhomogeneity of the plasma $\hat{\alpha}_{S3}$, and its small correction $\hat{\alpha}_{S4}$. Nevertheless, when the boosted Kerr metric is considered, three more terms appear: $\hat{\alpha}_{B1}$, $\hat{\alpha}_{B2}$, and $\hat{\alpha}_{B3}$. These are contributions due to the dragging of the inertial frame. The former is a constant that appears in all models considered, while the others two depend on the plasma distribution.\\

From now on, let us suppose that the vectors $\vec{J}_r$ and $\vec{b}$ are perpendicular to each other ($\cos\chi=0$). Therefore, the contribution $\hat{\alpha}_\chi$ vanishes (see Eq.~(\ref{chi})) and $\sin\chi=1$. Furthermore, since $\hat{\alpha}_{S4}$ is small, we neglect its contribution (see \cite{Kogan10}).
   
\subsection{\label{sec:singular} Singular isothermal sphere}
In this subsection, we consider the model for a singular isothermal sphere proposed in~\cite{Chandreaskahr39a,Binney87}. In this model, often used in lens modelling of galaxies and clusters, the density distribution has the form 
\begin{equation}
\label{density_distribution}
\rho(r)=\frac{\sigma^2_v}{2\pi r^2} ,
\end{equation} 
where $\sigma^2_v$ is a one-dimensional velocity dispersion. The concentration of the plasma has the form
\begin{equation}
\label{concentration_plasma}
N(r)=\frac{\rho(r)}{\kappa m_p}, 
\end{equation}
where  $m_p$ is the proton mass and $\kappa$ is a non-dimensional coefficient which is related to the dark matter contribution~\cite{Kogan10}. Using Eqs.~(\ref{refraction_index_inhomogeneous_plasma}) and (\ref{density_distribution}) the plasma frequency is 
\begin{equation}
\label{plasma_frequency_SIS}
\omega^2_e=K_eN(r)=\frac{K_e\sigma^2_v}{2\pi\kappa m_pr^2}.
\end{equation}
Then, from Eqs.~(\ref{non_uniform_plasma}) and (\ref{plasma_frequency_SIS}), and the well known property of the $\Gamma$-function \cite{Gradshteyn07} (see Appendix II), the contributions to the deflection angle can be found in the form
\begin{equation}
\label{contributions_SIS}
\begin{array}{cc}
\hat{\alpha}_{S2}=\frac{1}{12\pi}\frac{\omega^2_c}{\omega^2\overline{b}^3},&
\hat{\alpha}_{S3}=-\frac{1}{16}\frac{\omega^2_c}{\omega^2\overline{b}^2}\\\\
\hat{\alpha}_{B2}=-\frac{1}{48\pi}\frac{\widetilde{J}_r\omega^2_c}{\Lambda \omega^2\overline{b}^4},&\hat{\alpha}_{B3}=\frac{1}{20\pi}\frac{\widetilde{J}_r\omega^2_c}{\Lambda\omega^2\overline{b}^4}.\\
\end{array}
\end{equation}
Where $\omega^2_c=\frac{K_e\sigma^2_v}{M^2\kappa m_p}$, $\widetilde{J}_r=J_r/M^2$, and $\overline{b}=b/2M$. Hence deflection angle takes the form
\begin{eqnarray}
\label{alpha_SIS}
\hat{\alpha}_{SIS}&=&\frac{2}{\overline{b}}+\frac{1}{12\pi}\frac{\omega^2_c}{\omega^2\overline{b}^3}-\frac{1}{16}\frac{\omega^2_c}{\omega^2\overline{b}^2}
+\frac{1}{2}\frac{\widetilde{J}_r}{\Lambda \overline{b}^2}\nonumber \\
&&-\frac{1}{48\pi}\frac{\widetilde{J}_r\omega^2_c}{\Lambda \omega^2\overline{b}^4}+\frac{1}{20\pi}\frac{\widetilde{J}_r\omega^2_c}{\Lambda\omega^2\overline{b}^4}
\end{eqnarray}

\begin{figure}[h!]
\includegraphics[scale=0.38]{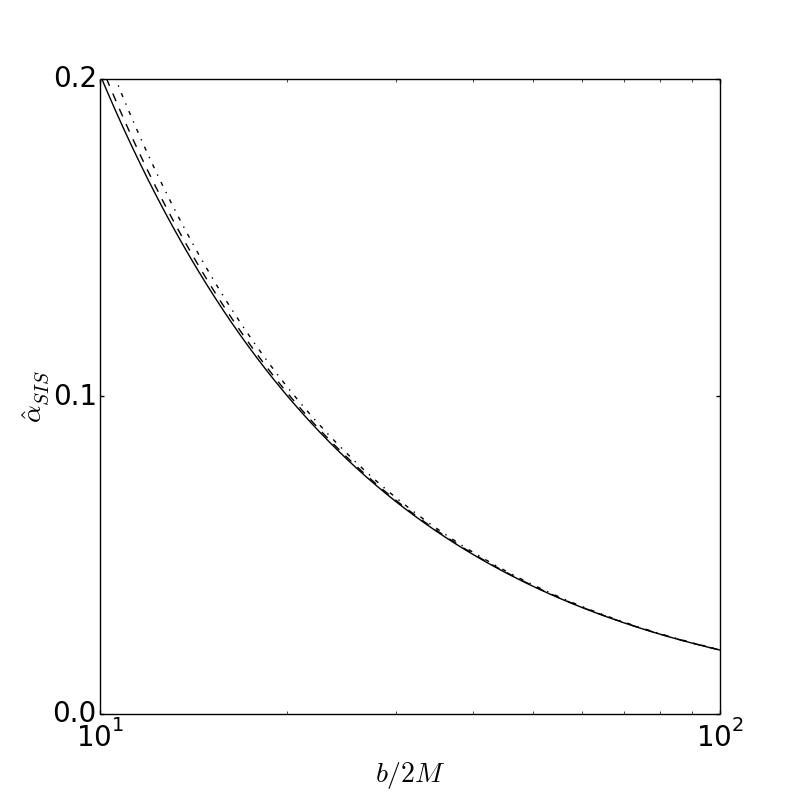}
\label{alpha_vs_b2M_uniform_plasma_rotating_case}
\caption{Plot of $\hat{\alpha}_{SIS}$ vs. $b/2M$ for $\Lambda=1$ (continuous line), $\Lambda=0.2$ (dashed line), and $\Lambda=0.1$ (dot-dashed line). We used $J_r/M^2=0.25$, $\sin\chi=1$, and $\omega^2_c/\omega^2=0.5$. \label{fig6}}
\end{figure}
\begin{figure}[h!]
\includegraphics[scale=0.38]{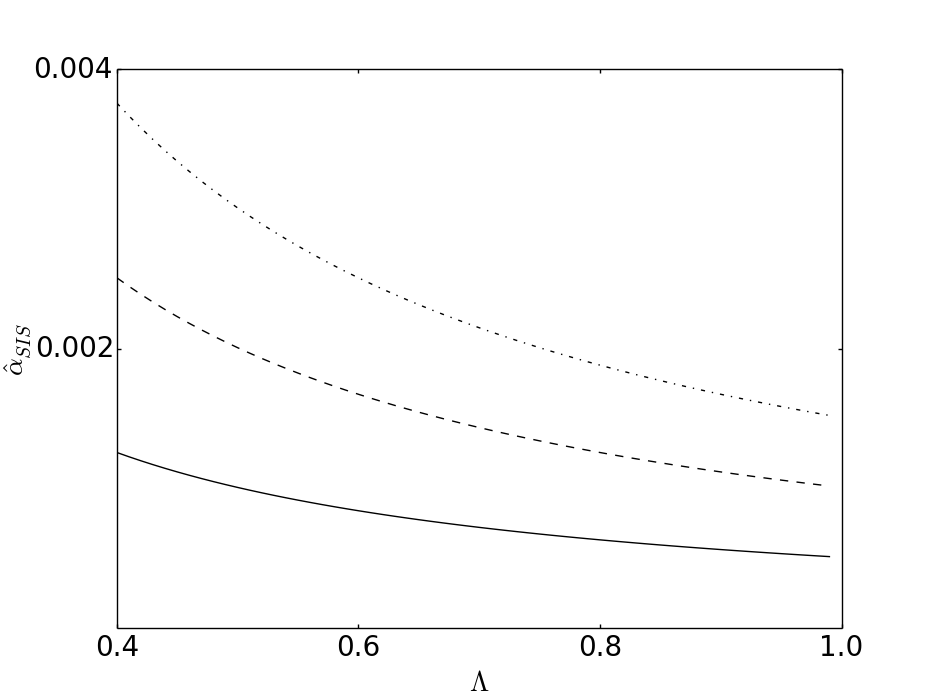}
\label{alpha_vs_b2M_uniform_plasma_rotating_case}
\caption{Plot of $\hat{\alpha}_{SIS}$ vs. $\Lambda$ for $\widetilde{J}_r=0.1$ (continuous line), $\widetilde{J}_r=0.2$ (dashed line), and $\widetilde{J}=0.3$ (dot-dashed line). We used, $\overline{b}=10$, $\sin\chi=1$, and $\omega^2_c/\omega^2=0.5$. \label{fig7}}
\end{figure}

In Fig.~\ref{fig6}, we plot $\hat{\alpha}_{SIS}$ as a function of $\overline{b}$ for different values of $\Lambda$. The figure does not show any difference for values of $b/2M$ greater than 10. However, for values of $b/2M$ near to 10, we see a small difference. This means that $\hat{\alpha}_{SIS}$ is greater when $\Lambda$ is small. For $\Lambda=1$ ($v=0$), we have the case of a slowly rotating massive object. Therefore, the parameter $\Lambda$ has a small effect on the deflection angle. This tendency can be seen clearly in Fig.~\ref{fig7}, where we plotted the behavior of the deflection angle as a function of $\Lambda$ for different values of $\widetilde{J}_r$. Note that the boosted parameter is constrained to be in the interval $0<\Lambda\leq 1$. \\

\begin{figure}[h!]
\includegraphics[scale=0.38]{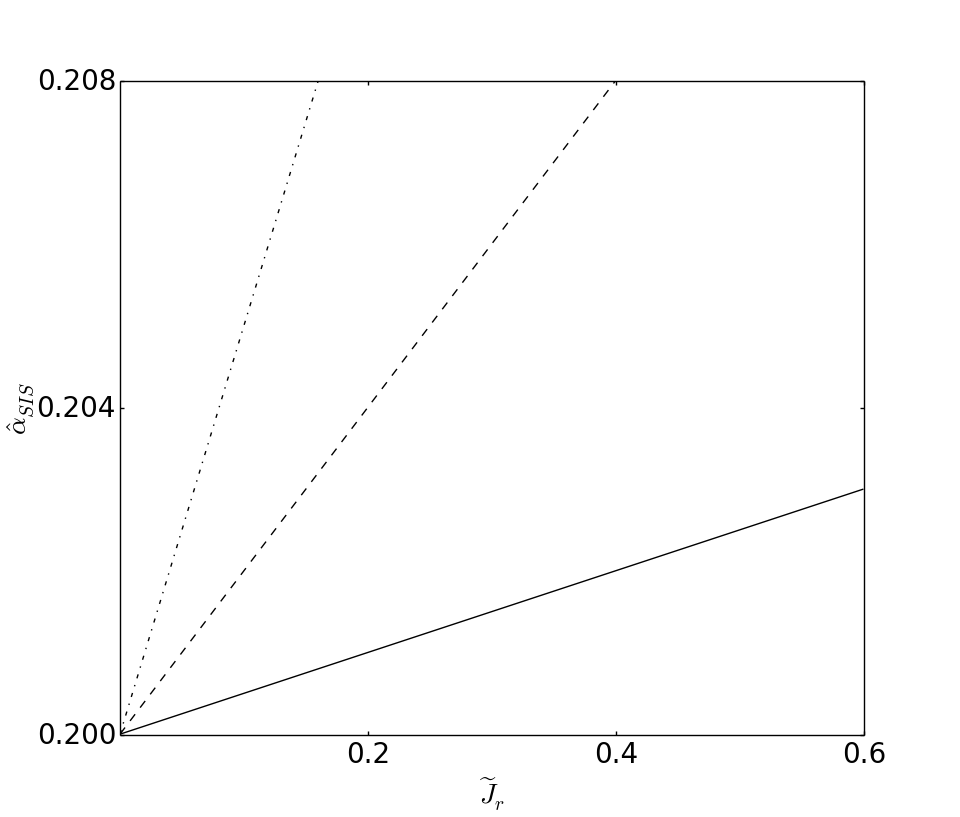}
\label{alpha_vs_b2M_uniform_plasma_rotating_case}
\caption{Plot of $\hat{\alpha}_{SIS}$ vs. $\widetilde{J}_r$ for $\Lambda=1$ (continuous line), $\Lambda=0.25$ (dashed line), and $\Lambda=0.1$ (dot-dashed line). We used, $\overline{b}=10$, $\sin\chi=1$, and $\omega^2_c/\omega^2=0.5$. \label{fig8}}
\end{figure}
In Fig.~\ref{fig8}, on the other hand, we plot $\hat{\alpha}_{SIS}$ as a function of $\widetilde{J}_r$ for different values of $\Lambda$. From this figure we conclude that, not only the dragging of the inertial system, but also the boosted parameter $\Lambda$ contribute to the deflection angle: the greater the values of $\widetilde{J}_r$ (plus small values of $\Lambda$) the greater the value of the deflection angle $\hat{\alpha}_{SIS}$.

\subsection{\label{sec:nonsingular} Non-singular isothermal gas sphere}
Now we consider a gravitational lens model for an isothermal sphere. For this model, the singularity at the origin is replaced by a finite core and the density distribution is given in~\cite{Hinshaw87} 

\begin{eqnarray}
\label{Density_distribution_nosingular}
\rho(r)=\frac{\sigma^2_v}{2\pi(r^2+r^2_c)}=\frac{\rho_0}{\left(1+\frac{r^2}{r^2_c}\right)},&&\rho_0=\frac{\sigma^2_v}{2\pi r^2_c},
\end{eqnarray}
where $r_c$ is the core radius.

Therefore, after substitution of Eq.~(\ref{Density_distribution_nosingular}) in Eqs.~(\ref{concentration_plasma}) and (\ref{plasma_frequency_SIS}), the plasma frequency is expressed as
\begin{equation}
\label{plasma_frequency_for_non_singular_isthermal_gas_sphere}
\omega^2_e=\frac{K_e\sigma^2_v}{2\pi\kappa m_p(r^2+r^2_c)}.
\end{equation}
Then, from Eqs.~(\ref{non_uniform_plasma}) and (\ref{plasma_frequency_SIS}), the contributions to the deflection angle are (see Appendix~II)
\begin{eqnarray}
\label{contributions_NSIS}
\hat{\alpha}_{S2}&=&\frac{2\overline{b}\omega^2_c}{\pi\omega^2}\bigg[\frac{1}{4\overline{b}^2\overline{r}^2_c}-\frac{{\rm arctanh}\bigg(\frac{\overline{r}_c}{\sqrt{4\overline{b}^2+\overline{r}^2_c}}\bigg)}{\overline{r}^3_c\sqrt{\overline{r}^2_c+4\overline{b}^2}}\bigg]\ ,\\
\hat{\alpha}_{S3}&=&-\frac{1}{2}\frac{\overline{b}\omega^2_c}{(4\overline{b}^2+\overline{r}^2_c)^\frac{3}{2}\omega^2}\ ,\\
\hat{\alpha}_{B2}&=&-\frac{\widetilde{J}_r\omega^2_c}{2\pi\Lambda\omega^2}\left[\frac{1}{4\overline{b}^2\overline{r}^2_c}-\frac{{\rm arctanh}\left(\frac{\overline{r}_c}{\sqrt{4\overline{b}^2+\overline{r}^2_c}}\right)}{\overline{r}^3_c\sqrt{\overline{r}^2_c+4\overline{b}^2}}\right], \\
\hat{\alpha}_{B3}&=&\frac{6}{\pi}\frac{\overline{b}^2\widetilde{J}_r\omega^2_c}{\Lambda\omega^2}\left[\frac{2\overline{r}^2_c-12\overline{b}^2}{48\overline{b}^4\overline{r}^4_c}+\frac{{\rm arctanh}\left(\frac{\overline{r}_c}{\sqrt{4\overline{b}^2+\overline{r}^2_c}}\right)}{\overline{r}^5_c\sqrt{\overline{r}^2_c+4\overline{b}^2}}\right],\nonumber\\
\end{eqnarray}
where $\omega^2_c=\frac{K_e\sigma^2_v}{M^2\kappa m_p}$, $\overline{r}_c=r_c/M$, $\widetilde{J}_r=J_r/M^2$, and $\overline{b}=b/2M$.\\

\begin{figure}[h!]
\includegraphics[scale=0.38]{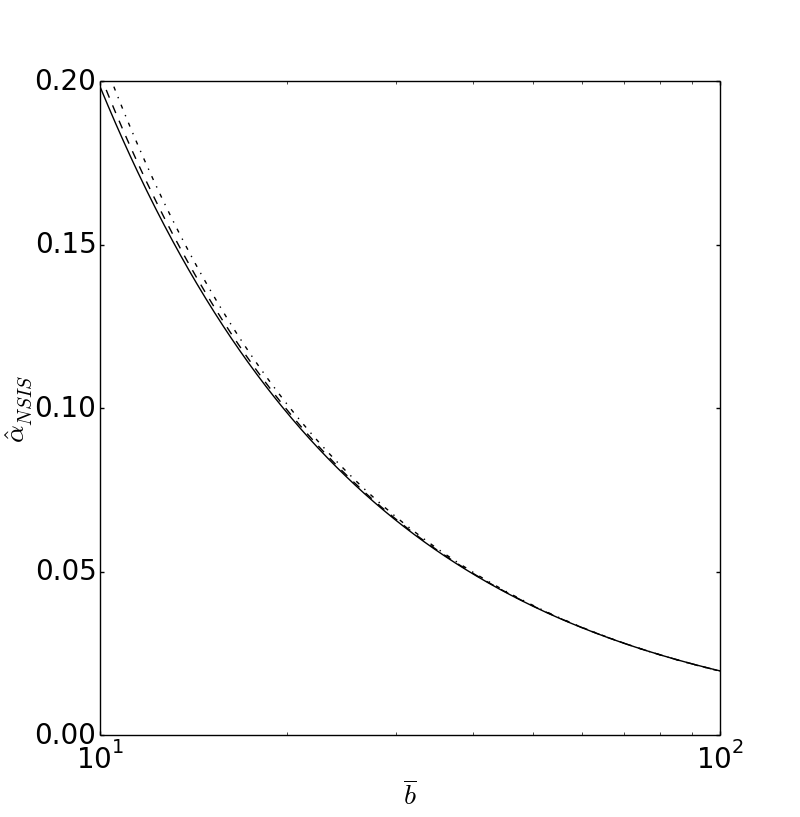}
\label{alpha_vs_b2M_uniform_plasma_rotating_case}
\caption{Plot of $\hat{\alpha}_{NSIS}$ vs. $\overline{b}$ for $\Lambda=1$ (continuous line), $\Lambda=0.25$ (dashed line), and $\Lambda=0.1$ (dot-dashed line). We used, $\widetilde{J}_r=0.25$, $\overline{r}_c=10$, $\sin\chi=1$, and $\omega^2_c/\omega^2=0.5$.\label{fig9}}
\end{figure}

\begin{figure}[h!]
\includegraphics[scale=0.38]{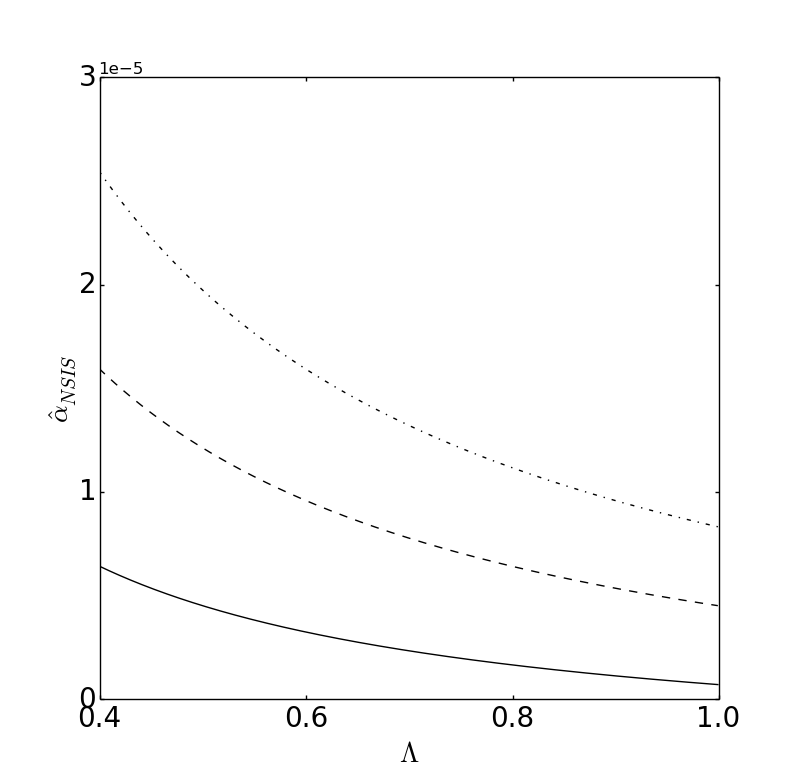}
\label{alpha_vs_b2M_uniform_plasma_rotating_case}
\caption{Plot of $\hat{\alpha}_{NSIS}$ vs. $\Lambda$ for $\widetilde{J}_r=0.1$ (continuous line), $\widetilde{J}_r=0.2$ (dashed line), and $\widetilde{J}=0.3$ (dot-dashed line). We used, $\overline{b}=100$, $\overline{r}_c=10$, $\sin\chi=1$, and $\omega^2_c/\omega^2=0.5$. Note the scale used for the deflection angle: each value is multiplied by $1e-5=1\times 10^{-5}$ \label{fig10}}
\end{figure}

In Fig.~\ref{fig9} we plot $\hat{\alpha}_{NSIS}$ as a function of $\overline{b}$ for different values of $\Lambda$. In the plot, we have $\overline{b} \gg \overline{r}_c$ because we are in the weak field limit. According to the figure, the behavior is quite similar to that of the deflection angle in the case of a singular plasma distribution: there are small differences in $\hat{\alpha}_{NSIS}$ when small values of $\Lambda$ are considered, and no there is no difference in the deflection angle when the impact parameter $\overline{b}$ takes values greater than $10$. Fig.~\ref{fig10} helps to see this behavior clearly.\\

\begin{figure}[h!]
\includegraphics[scale=0.38]{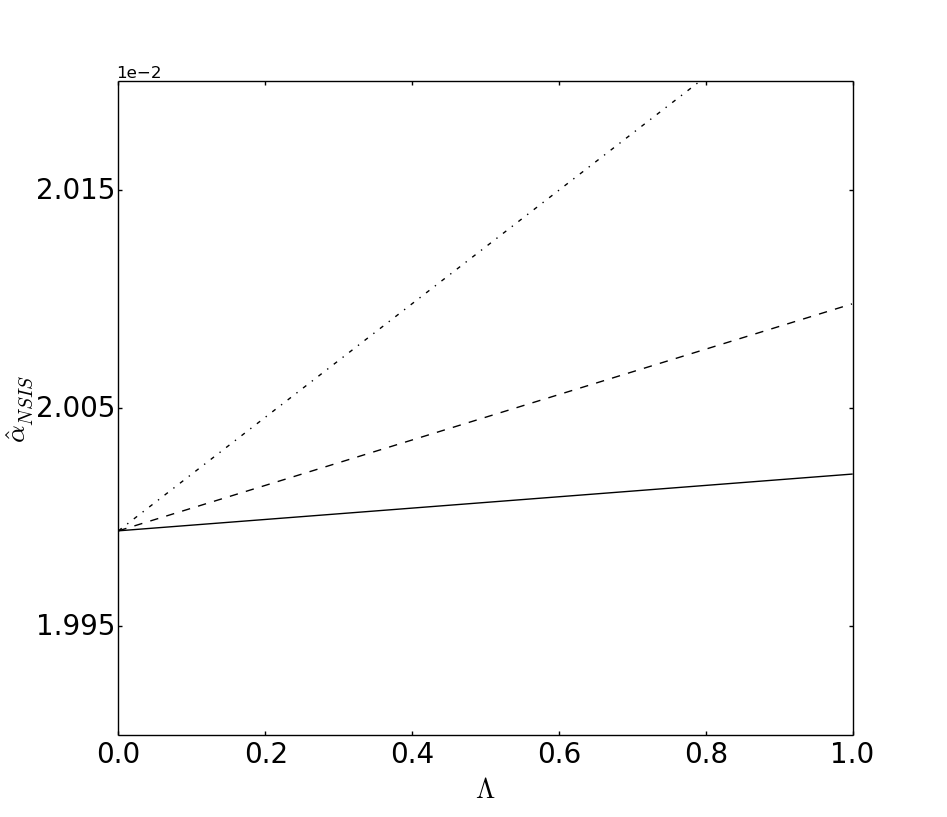}
\label{alpha_vs_b2M_uniform_plasma_rotating_case}
\caption{Plot of $\hat{\alpha}_{NSIS}$ vs. $\widetilde{J}_r$ for $\Lambda=1$ (continuous line), $\Lambda=0.25$ (dashed line), and $\Lambda=0.1$ (dot-dashed line). We used, $\overline{b}=100$, $\overline{r}_c=10$, $\sin\chi=1$, and $\omega^2_c/\omega^2=0.5$. Note the scale used for the deflection angle: each value is multiplied by $1e-2=1\times 10^{-2}$ \label{fig11}}
\end{figure}

In Fig.~\ref{fig11} we plot the deflection angle as a function of $\widetilde{J}_r$ for different values of $\Lambda$. Once again, the dragging of the inertial system along with small values of the boosted parameter $\Lambda$ play an important role when compared with the slowly rotating case~\cite{Morozova13}.

\subsection{ \label{sec:galaxy_cluster}Plasma in a galaxy cluster}
In a galaxy cluster, due to the large temperature of the electrons, the distribution of electrons may be homogeneous. Therefore, it is proper to suppose a singular isothermal sphere as a model for the distribution of the gravitating matter. Using this approximation, and without considering the mass of the plasma, Bisnovatyi-Kogan and O. Yu. Tsupko solved the equation of hydrostatic equilibrium of a plasma in a gravitational field finding that the plasma density distribution has the form \cite{Kogan10}.
\begin{equation}
\label{plasma_distribution_in_cluster}
\rho(r)=\rho_0\left(\frac{r}{r_0}\right)^{-s},s=\frac{2\sigma^2_v}{\mathfrak{R}T},
\end{equation}   
and the plasma frequency is equal to 
\begin{equation}
\label{plasma_frequency_cluster}
\omega^2_e=\frac{\rho_0K_e}{\kappa m_p}\left(\frac{r}{r_0}\right)^{-s}.
\end{equation}
Hence, using Eqs.~(\ref{non_uniform_plasma}) and (\ref{plasma_frequency_SIS}) once again, the contributions to the deflection angle are (see Appendix~II)
\begin{eqnarray}
\label{contributions_PGC}
\hat{\alpha}_{S2}&=&\frac{\sqrt{\pi}}{2^{s+1}(s+1)}\frac{\overline{r}^s_0\omega^2_f}{\overline{b}^2\omega^2}\frac{\Gamma(\frac{s}{2}+1)}{\Gamma(\frac{s+1}{2})}\ ,\\
\hat{\alpha}_{S3}&=&-\frac{\sqrt{\pi}}{2^s}\frac{\omega^2_f}{\omega^2}\frac{\Gamma(\frac{s}{2}+1)}{\Gamma(\frac{s}{2})}\left(\frac{\overline{r}_0}{\overline{b}}\right)^s\ ,\\
\hat{\alpha}_{B2}&=&-\frac{\pi}{2^{s+2}(s+1)}\frac{\widetilde{J}_r\overline{r}^2_0\omega^2_f}{\overline{b}^{s+2}\Lambda\omega^2}\frac{\Gamma(\frac{s}{2}+1)}{\Gamma(\frac{s+1}{2})}\ ,\\
\hat{\alpha}_{B3}&=&\frac{3\sqrt{\pi}}{2^{s+2}(s+3)}\frac{\widetilde{J}_r \overline{r}^s_0\omega^2_f}{b^{s+2}\Lambda\omega^2}\frac{\Gamma(\frac{s+4}{2})}{\Gamma(\frac{s+1}{2})}\ ,
\end{eqnarray}
where $\omega^2_f=\frac{K_e\rho_0}{\kappa m_p}$, $\overline{r}_0=r_0/M$, $\widetilde{J}_r=J_r/M^2$, and $\overline{b}=b/2M$.\\

\begin{figure}[h!]
\includegraphics[scale=0.38]{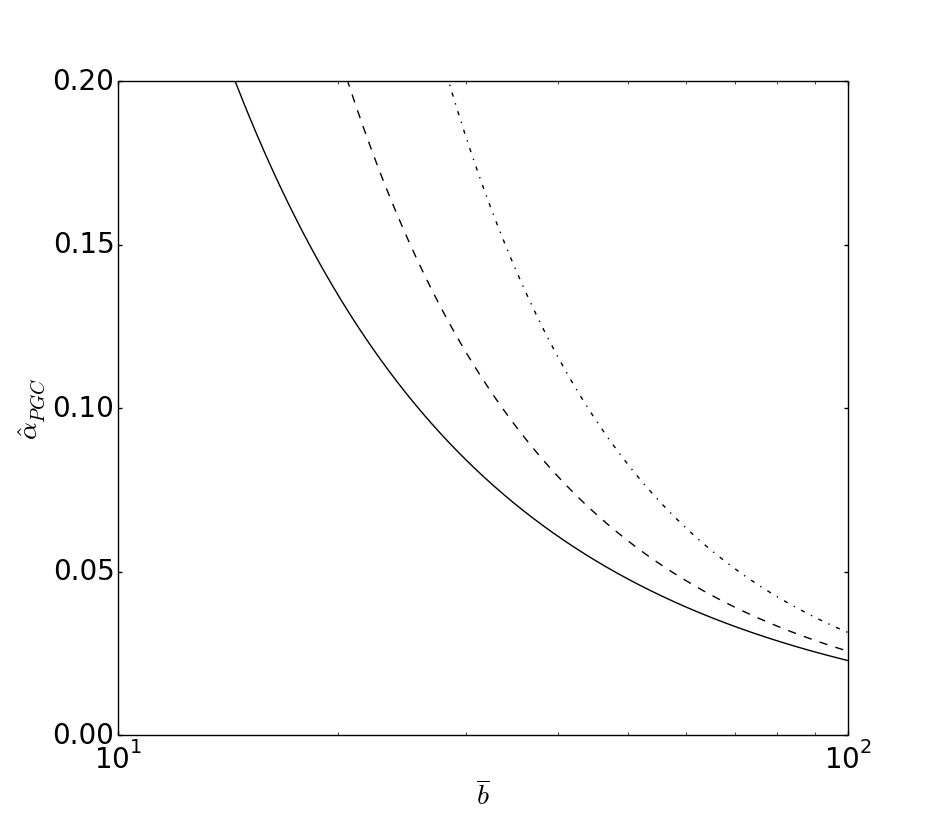}
\label{alpha_vs_b2M_uniform_plasma_rotating_case}
\caption{Plot of $\hat{\alpha}_{PGC}$ vs. $\overline{b}$ for $\Lambda=1$ (continuous line), $\Lambda=0.25$ (dashed line), and $\Lambda=0.1$ (dot-dashed line). We used, $\widetilde{J}_r=0.25$, $\overline{r}_0=10$, $\sin\chi=1$, $s=0.03$, and $\omega^2_f/\omega^2=0.5$.\label{fig12}}
\end{figure}

\begin{figure}[h!]
\includegraphics[scale=0.38]{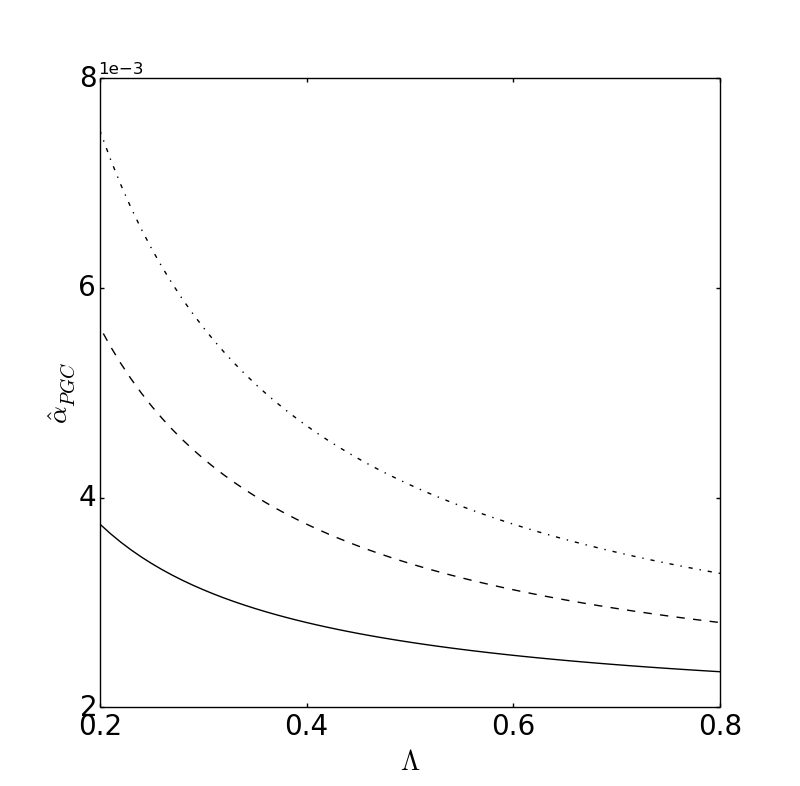}
\label{alpha_vs_b2M_uniform_plasma_rotating_case}
\caption{Plot of $\hat{\alpha}_{PGC}$ vs. $\Lambda$ for $\widetilde{J}_r=0.1$ (continuous line), $\widetilde{J}_r=0.2$ (dashed line), and $\widetilde{J}_r=0.3$ (dot-dashed line). We used $\overline{r}_0=10$, $\sin\chi=1$, $s=0.03$, $\overline{b}=100$ and $\omega^2_f/\omega^2=0.5$. Note the scale used for the deflection angle: each value is multiplied by $1e-3=1\times 10^{-3}$\label{fig13}}
\end{figure}

\begin{figure}[h!]
\includegraphics[scale=0.38]{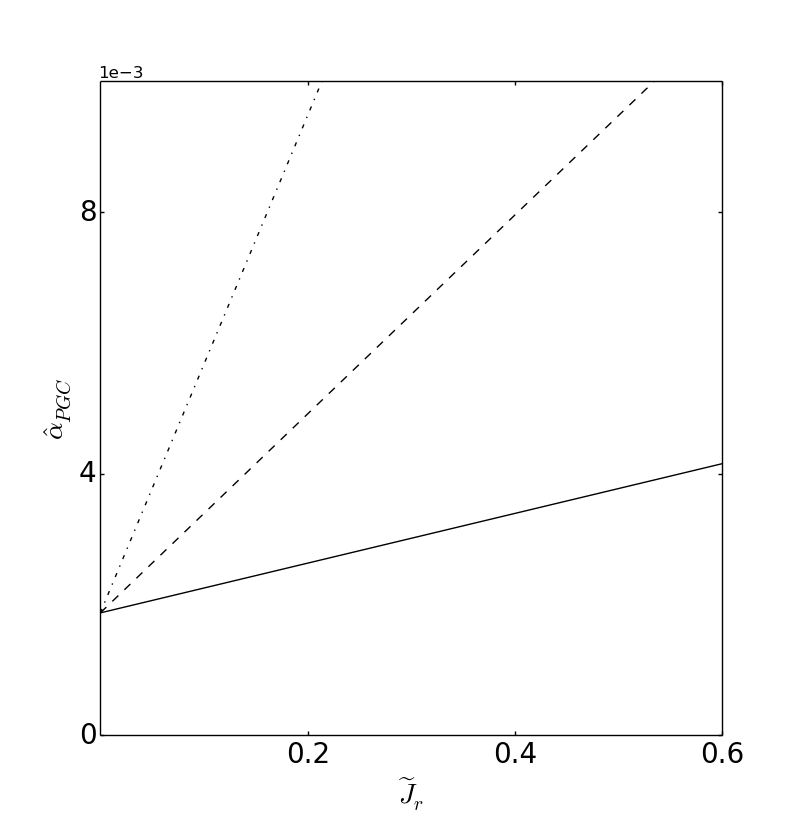}
\caption{Plot of $\hat{\alpha}_{PGC}$ vs. $\widetilde{J}_r$ for $\Lambda=1$  (continuous line), $\Lambda=0.25$ (dashed line), and $\Lambda=0.1$ (dot-dashed line). We used $\overline{r}_0=1.2$, $\sin\chi=1$, $s=0.03$, $\overline{b}=100$ and $\omega^2_f/\omega^2=0.5$. Note the scale used for the deflection angle: each value is multiplied by $1e-3=1\times 10^{-3}$\label{fig14}}
\end{figure}

In Figs.~\ref{fig12}, \ref{fig13}, and \ref{fig14} we plot $\hat{\alpha}_{PGC}$ as a function of $\overline{b}$, $\Lambda$, and $\widetilde{J}_r$, respectively. In order to obtain these plots we considered the case $s<<1$ \cite{Kogan10}. According to Figs.~\ref{fig12} and \ref{fig13}, differences in the deflection angle can be seen clearly for the \textbf{PGC} distribution when compared with the previous distributions. Furthermore, Fig.~\ref{fig14} shows that the deflection angle increases due to the dragging and small values of $\Lambda$.\\

\begin{figure}[h!]
\includegraphics[scale=0.38]{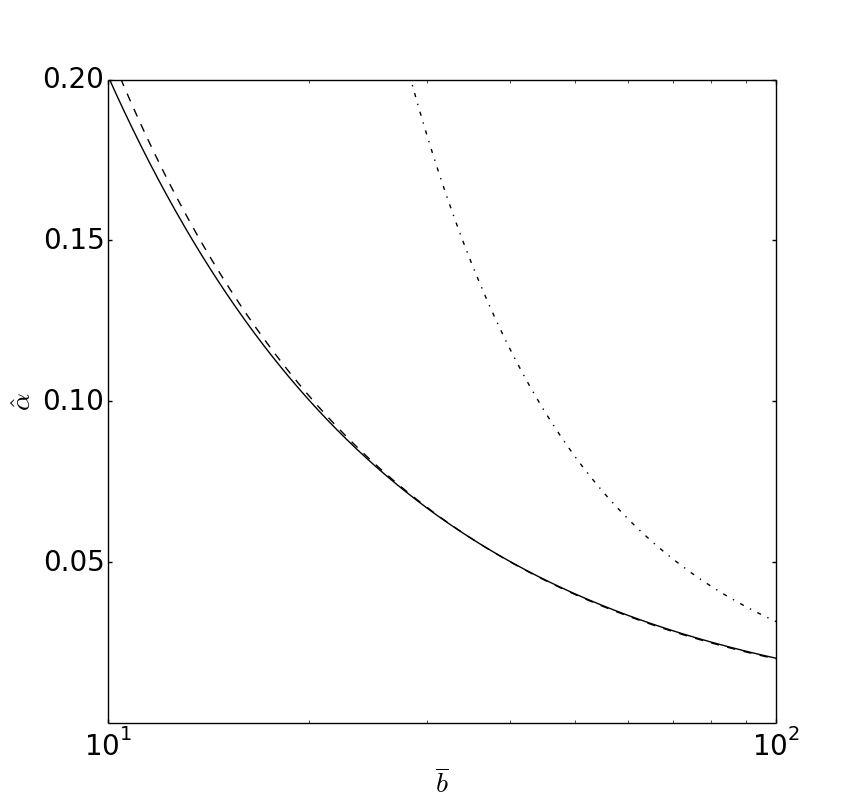}
\label{alpha_vs_b2M_uniform_plasma_rotating_case}
\caption{Plot of $\hat{\alpha}$ vs. $\overline{b}$ for  \textbf{SIS} (continuous line), \textbf{NSIS} (dashed line), and \textbf{PGC} (dot-dashed line). We used $\Lambda=0.1$, $\overline{r}_c=10$, $\overline{r}_0=10$, $\sin\chi=1$, $s=0.03$, and $\omega^2_f/\omega^2=\omega^2_c/\omega^2=0.5$. For \textbf{NSIS} we use $\Lambda=1$ since no difference from \textbf{SIS} was found.\label{fig15}}
\end{figure}

On the other hand, in Fig.~\ref{fig15}, we plotted the behavior of the deflection angle for all distributions as a function of the impact parameter $\overline{b}$. Note that the values of $\hat{\alpha}$ for the \textbf{PGC} distribution are grater than the other two distributions. In the figure there is a small difference between \textbf{SIS} and \textbf{NSIS} distributions for small values of $b/2M$.\\  

\begin{figure}[h!]
\includegraphics[scale=0.38]{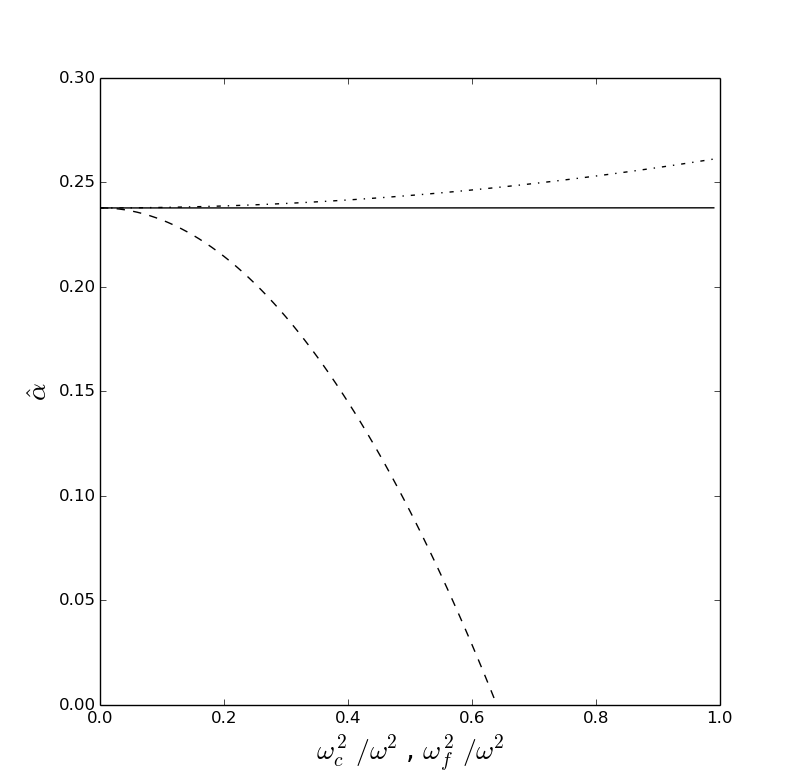}
\caption{Plot of $\hat{\alpha}$ vs. $\omega^2_f/\omega^2$, $\omega^2_c/\omega^2$  for  \textbf{SIS} (continuous line), \textbf{NSIS} (dashed line), and \textbf{PGC} (dot-dashed line). We used $\Lambda=0.1$, $\overline{r}_c=10$, $\overline{r}_0=10$, $\sin\chi=1$, $s=0.03$, and $\overline{b}=9$\label{fig16}}
\end{figure}

Finally, in Fig.~\ref{fig16} we plotted $\hat{\alpha}$ as a function of $\omega^2_c/\omega^2$ (for \textbf{SIS} and \textbf{NSIS}) and $\omega^2_f/\omega^2$ (for \textbf{PGC}). This figure clearly show that the deflection angle is more affected by the plasma for the \textbf{PGC} distribution than the other two for values of $\omega^2_f/\omega^2$ greater than $0.4$.

\section{\label{sec:magnification} Lens equation and magnification in the presence of plasma}
In this section, we compute the magnification for the boosted Kerr metric in the presence of plasma. We consider the uniform and the \textbf{SIS} plasma distributions discussed previously in sections \ref{sec:Deflection_angle_for_the_slowly_rotating_case} and \ref{sec:Models} respectively.\\

The magnification of brightness of the star is defined by the relation\cite{Morozova13}
\begin{equation}
\label{Magnification}
\begin{array}{cc}
\mu_\Sigma=\frac{I_{tot}}{I_{\ast}}=\sum_k \left|\left(\frac{\theta_k}{\beta}\right)\left(\frac{d\theta_k}{d\beta}\right)\right|,&k=1,2,...,m,
\end{array}
\end{equation}
where $m$ is the number of images, $I_{tot}$ is the total brightness of the images, $I_{\ast}$ is the unlensed brightness of the source, $\theta_k$ is the position of the image, and $\beta$ is the angular position of the source (see figure \ref{esquema}). In this sense, in order to compute the contribution of the boosted parameter $\Lambda$ to magnification, we have to solve the lens equation; which is given by the relation\cite{Morozova13}
\begin{equation}
\label{lens_equation}
\theta D_s=\beta D_s+\hat{\alpha}D_{ls},
\end{equation} 
here $D_s$ is the distance from the observer to the source, $D_{ls}$ is the distance from the lens to the source, $\hat{\alpha}$ is the deflection angle, and $\theta$, $\beta$ the positions of the image and the source respectively (see figure \ref{esquema}). 
\subsection{Uniform plasma}
In the case of small angles, it is well known that the impact parameter can be expressed as 
\begin{equation}
\label{small_angles}
b\approx D_l\theta,
\end{equation}
where $D_l$ is the distance from the observer to the lens. Therefore, after using equation (\ref{deflection_angle_rotating_case_homgeneous_plasma}), the lens equation for the slowly rotating case in the presence of uniform plasma takes the form
\begin{equation}
\label{Lens_slowly_rotating}
\theta^3-\beta\theta^2-\frac{\theta^2_E}{2}\left(1+\frac{1}{1-\frac{\omega^2_e}{\omega^2}}\right)\theta-\frac{\theta^2_E\tilde{J}_r}{4\overline{D}_l\Lambda}\frac{1}{\sqrt{1-\frac{\omega^2_e}{\omega^2}}}=0.
\end{equation}
In the last expression, in order to be consistent with the notation, we use $\overline{b}\approx\overline{D}_l\theta$, where $\overline{D}_l=D_l/2M$. Furthermore, we have defined
\begin{equation}
\label{Einstein_ring}
\theta^2_E=\frac{4MD_{ls}}{D_{l}D_{s}}=\frac{2\overline{D}_{ls}}{\overline{
D}_l\overline{D}_s},
\end{equation}
with $\overline{D}_{ls}=D_{ls}/2M$ and $\overline{D}_s=D_s/2M$. $\theta_E$ is known as the Einstein angle. Note that equation (\ref{Lens_slowly_rotating}) reduces to that obtained by \cite{Morozova13} for $\Lambda=1$ ($v=0$).\\

In order to solve equation (\ref{Lens_slowly_rotating}) we introduce a new variable $x$ by the relation (see \cite{Morozova13,Abramowitz72} for details.)
\begin{equation}
\label{new_variable}
\theta=x+\frac{\beta}{3};
\end{equation}   
form which equation (\ref{Lens_slowly_rotating}) reduces to
\begin{equation}
\label{new_lens_equation_rotating_case}
x^3+px+q=0,
\end{equation}
where
\begin{equation}
\label{p_and_q}
\begin{aligned}
p&=-\frac{\beta^2}{3}-\frac{\theta^2_E}{2}\left(1+\frac{1}{1-\frac{\omega^2_e}{\omega^2}}\right)\\
q&=-\frac{2\beta^3}{27}-\frac{\beta\theta^2_E}{6}\left(1+\frac{1}{1-\frac{\omega^2_e}{\omega^2}}\right)-\frac{\theta^2_E\tilde{J}_r}{4\overline{D}_l\Lambda}\frac{1}{\sqrt{1-\frac{\omega^2_e}{\omega^2}}}.\\
\end{aligned}
\end{equation}
Note that the variable $q$, in contrast with the result obtained by \cite{Morozova13}, depends on the boosted parameter $\Lambda$.\\
 
Equation (\ref{new_lens_equation_rotating_case}) has three different real roots if  
\begin{equation}
\frac{q^2}{4}+\frac{p^3}{27}<0.
\end{equation}
Therefore, the solution has the form 
\begin{equation}
\begin{array}{cc}
x=2\sqrt[3]{r}\cos\frac{\phi+2k\pi}{3},&\hspace{1cm}k=0,1,2
\end{array}
\end{equation}
with
\begin{equation}
\begin{array}{cc}
r=\sqrt{-\frac{p^3}{27}},\hspace{1cm}&\cos\phi=-\frac{q}{2r}.
\end{array}
\end{equation}
Hence, after using equations (\ref{Magnification}) and (\ref{new_variable}), we obtain
\begin{equation}
\label{magnification_uniform_plasma}
\begin{aligned}
\mu_{\Sigma tot}&=\sum_k \left|\frac{\theta_k}{\beta} \frac{d\theta_k}{d\beta}\right|=\sum_k \left|\frac{x_k+\beta/3}{\beta}\left(\frac{dx_k}{d\beta}+\frac{1}{3}\right)\right|\\
&=\sum_k \left|\frac{1}{3\beta}\left(2\sqrt[3]{r}\cos\frac{\phi+2k\pi}{3}+\frac{\beta}{3}\right)\right|\\
&\times\left|\left[\frac{2r_\beta}{\sqrt[3]{r^2}}\cos\frac{\phi+2k\pi}{3}-2
\sqrt[3]{r}\phi_\beta\sin\frac{\phi+2k\pi}{3}+1\right]\right|
\end{aligned}
\end{equation}
for $k=0,1,2$. The subscript $\beta$ denotes the derivatives of the corresponding variables with respect to $\beta$.\\

In reference \cite{Bisnovatyi2010} the authors found that the magnification for small values of $\beta$ has the form (see equation (32) in \cite{Morozova13})\\
\begin{equation}
\label{Magnification_Bisnov}
\mu=\frac{1}{2}\frac{\sqrt{2\theta^2_E\left(1+\frac{1}{1-\frac{\omega^2_e}{\omega^2}}\right)}}{\beta}.
\end{equation}
Therefore, in order to study the behaviour of the magnification for small values of $\beta$, and compare with the case of uniform plasma studied by Bisnovatyi-Kogan and Tsupko (2010), it is necessary to express equation (\ref{magnification_uniform_plasma}) in the limit $\beta\rightarrow 0$. Hence, for small values of $\beta$ we have that
\begin{equation}
\label{beta_zero}
\begin{aligned}
\sqrt[3]{r}&\rightarrow\sqrt{\frac{1}{6}\theta^2_E\left(1+\frac{1}{1-\frac{\omega^2_e}{\omega^2}}\right)},\\
r_\beta&\rightarrow 0,\\
\phi_\beta&\rightarrow\frac{1}{\sqrt{1-\left(\frac{q}{2r}\right)^2}}\frac{q_\beta}{2r}\\
q_\beta&\rightarrow-\frac{\beta\theta^2_E}{6}\left(1+\frac{1}{1-\frac{\omega^2_e}{\omega^2}}\right)\\
\cos\phi&=-\frac{q}{2r}\rightarrow\frac{\sqrt{27}}{\sqrt{2}}\frac{1}{\theta_E}\frac{\tilde{J}_r}{\overline{D}_l\Lambda}\frac{1}{\sqrt{1-\frac{\omega^2_e}{\omega^2}}}\frac{1}{\sqrt{\left(1+\frac{1}{1-\frac{\omega^2_e}{\omega^2}}\right)^3}}.\\
\end{aligned}
\end{equation}
Where we have followed the same analysis done in \cite{Morozova13}. Note that $-q/2r$, in our case, depends on $\Lambda$. Thus, after using equations (\ref{magnification_uniform_plasma}), (\ref{Magnification_Bisnov}), and (\ref{beta_zero}), we found that $\mu_{\Sigma tot}/\mu$, in the limit $\beta\rightarrow0$, takes the form 
\begin{equation}
{\frac{\mu_{\Sigma tot}}{\mu}}=\frac{1}{3\sqrt{3}}\sum_{k}\left|\frac{\sin\frac{2(\phi+2k\pi)}{3}}{\sqrt{1-\left(\frac{q}{2r}\right)^2}}+2\cos\frac{\phi+2k\pi}{3}\right|.
\end{equation}
Now, setting $\tilde{J}_r=0$ and $\Lambda=1$, the last expression reduces to 
\begin{equation}
\label{matching_32_84}
\begin{aligned}
\frac{\mu_{\Sigma tot}}{\mu}&=\frac{1}{3\sqrt{3}}\sum_{k}\left|\sin\frac{(1+4k)\pi}{3}+2\cos\frac{(1+4k)\pi}{6}\right|\\
&=\frac{1}{3\sqrt{3}}\left| 2 \cos \left(\frac{\pi }{6}\right)+\sin \left(\frac{\pi }{3}\right)\right|\\
&+\frac{1}{3\sqrt{3}}\left| 2 \cos \left(\frac{5 \pi }{6}\right)+\sin \left(\frac{5 \pi }{3}\right)\right|\\ &+\frac{1}{3\sqrt{3}}\left| 2 \cos \left(\frac{9 \pi }{6}\right)+\sin (3 \pi )\right|=1.
\end{aligned}
\end{equation}
With this result, we have shown that the ratio $\mu_{\Sigma tot}/\mu$ is equal to unity when $\tilde{J}_r=0$ and $\Lambda=1$; this means that equation (\ref{magnification_uniform_plasma}) reduces to equation (\ref{Magnification_Bisnov}) in the limit $\beta\rightarrow 0$.\\

In Figs. \ref{fig17}\textcolor{blue}{.a} and \ref{fig17}\textcolor{blue}{.c}, we plotted the behaviour of the total magnification as a function of the boosted parameter $\Lambda$ for $\beta=0.001$ and $\beta=0.0001$ respectively. According to Fig. \ref{fig17}\textcolor{blue}{.a}, when $\beta=0.001$, the total magnification decreases as $\Lambda$ increases. This means that $\mu_{\Sigma tot}$ decreases as the boosted velocity $v$ of the black hole decreases. A similar behaviour can be seen from Fig. \ref{fig17}\textcolor{blue}{.c} when $\beta=0.0001$. Note that for small values of $\beta$, the magnitude of the total magnification increases. For example: when $\beta=0.001$ the total magnification is about $\mu_{\Sigma tot}\approx 52.2$. However, when $\beta=0.0001$, the value increases to $\mu_{\Sigma tot}\approx 522.2$.  

\subsection{Singular isothermal sphere}
In a similar way, in order to compute the magnification for \textbf{SIS}, we also use the approximation of small angles described in equation (\ref{small_angles}). Hence, after using equation (\ref{alpha_SIS}), the lens  equation for \textbf{SIS} takes the form
\begin{equation}
\label{lens_SIS}
\begin{aligned}
\theta^3-\beta\theta^2-\frac{2D_{ls}}{D_lD_s}\theta-\frac{\overline{D}_{ls}}{\overline{D}^2_l\overline{D}_s}\left(\frac{\tilde{J}_r}{2\Lambda}-\frac{\omega^2_c}{16\omega^2}\right)=0
\end{aligned}
\end{equation}   
In the last equation, as an approximation, we neglected the second and the last two terms of equation (\ref{alpha_SIS}) since they are very small in the weak field limit. Then, using equation (\ref{Einstein_ring}), equation (\ref{lens_SIS}) can be expressed in terms of the Einstein angle as: 
\begin{equation}
\label{lens_SIS_I}
\theta^3-\beta\theta^2-\theta^2_E\theta-\frac{\delta\theta^2_E}{\Lambda}=0,
\end{equation}   
where we defined:
\begin{equation}
\label{delta}
\begin{aligned}
\delta&=\frac{1}{\overline{D}_l}\left(\frac{\tilde{J}_r}{4}-\frac{\omega^2_c\Lambda}{32\omega^2}\right).\\
\end{aligned}
\end{equation}
Now, introducing the new variable $y=\theta+\beta/3$, the equation (\ref{lens_SIS_I}) reduces to 
\begin{equation}
\label{lens_SIS_new_varable}
y^3+my+n=0
\end{equation}
with,
\begin{equation}
\label{PQR}
\begin{aligned}
m=&-\frac{\beta^2}{3}-\theta^2_E\\
n=&-\frac{2\beta^3}{27}-\frac{\beta\theta^2_E}{3}-\frac{\delta\theta^2_E}{\Lambda}.\\
\end{aligned}
\end{equation}
Equation (\ref{lens_SIS_I}) has three different real roots if 
\begin{equation}
\frac{m^2}{4}+\frac{n^3}{27}<0.
\end{equation}
This condition is already satisfied in our case. Hence the solutions has the form 
\begin{equation}
\begin{array}{cc}
y=2\sqrt[3]{l}\cos\frac{\epsilon+2k\pi}{3},&k=0,1,2
\end{array}
\end{equation}
with
\begin{equation}
\begin{array}{cc}
l=\sqrt{-\frac{m^3}{27}},\hspace{1cm}&\cos\epsilon=-\frac{n}{2l}
\end{array}
\end{equation}
Therefore, after using equations (\ref{Magnification}) and the new variable $y$, we obtain
\begin{equation}
\label{magnification_SIS}
\begin{aligned}
\mu_\Sigma&=\sum_k \left|\frac{1}{3\beta}\left(2\sqrt[3]{l}\cos\frac{\epsilon+2k\pi}{3}+\frac{\beta}{3}\right)\right|\\
&\times\left|\left[\frac{2l_\beta}{\sqrt[3]{l^2}}\cos\frac{\epsilon+2k\pi}{3}-2
\sqrt[3]{l}\epsilon_\beta\sin\frac{\epsilon+2k\pi}{3}+1\right]\right|
\end{aligned}
\end{equation}
for $k=0,1,2$. The subscript $\beta$ has the same meaning as in equation (\ref{magnification_uniform_plasma}).\\

\begin{figure*}[t]
\begin{center}
a.\includegraphics[scale=0.65]{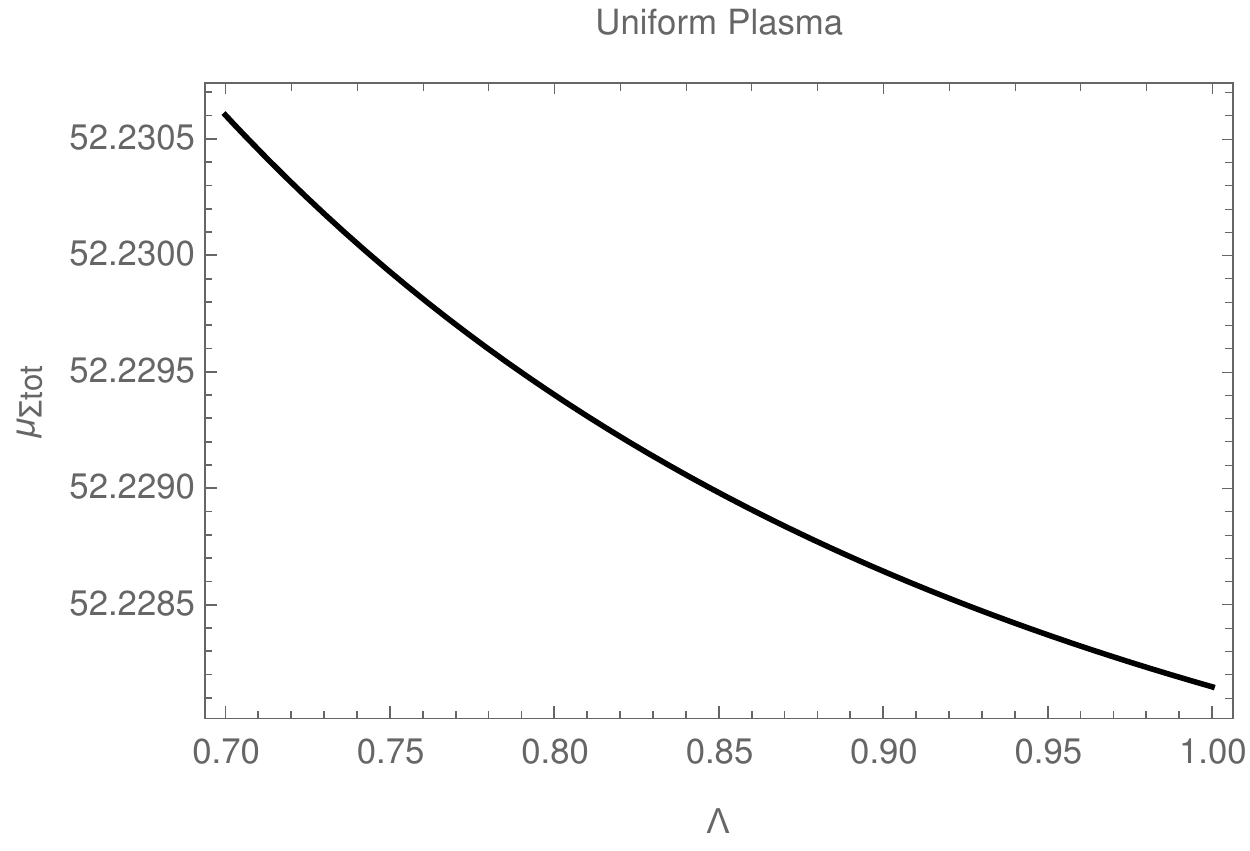}
b.\includegraphics[scale=0.65]{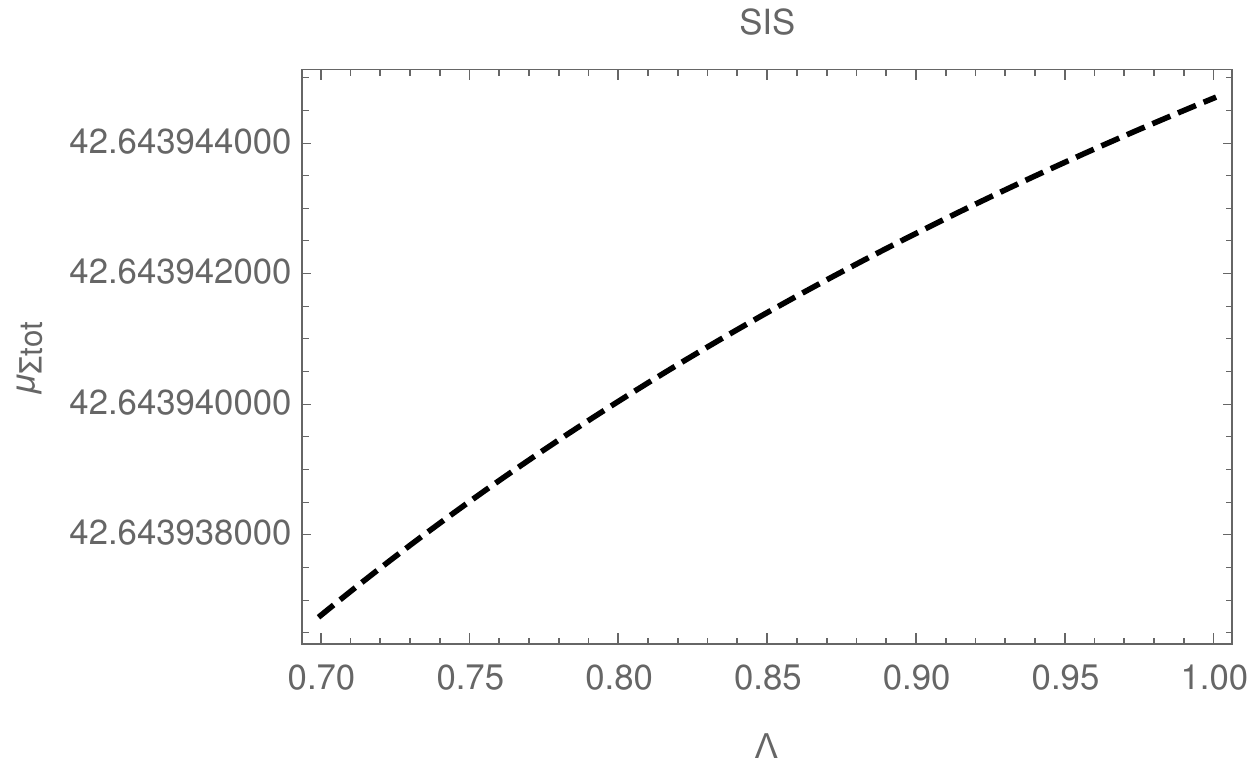}
c.\includegraphics[scale=0.65]{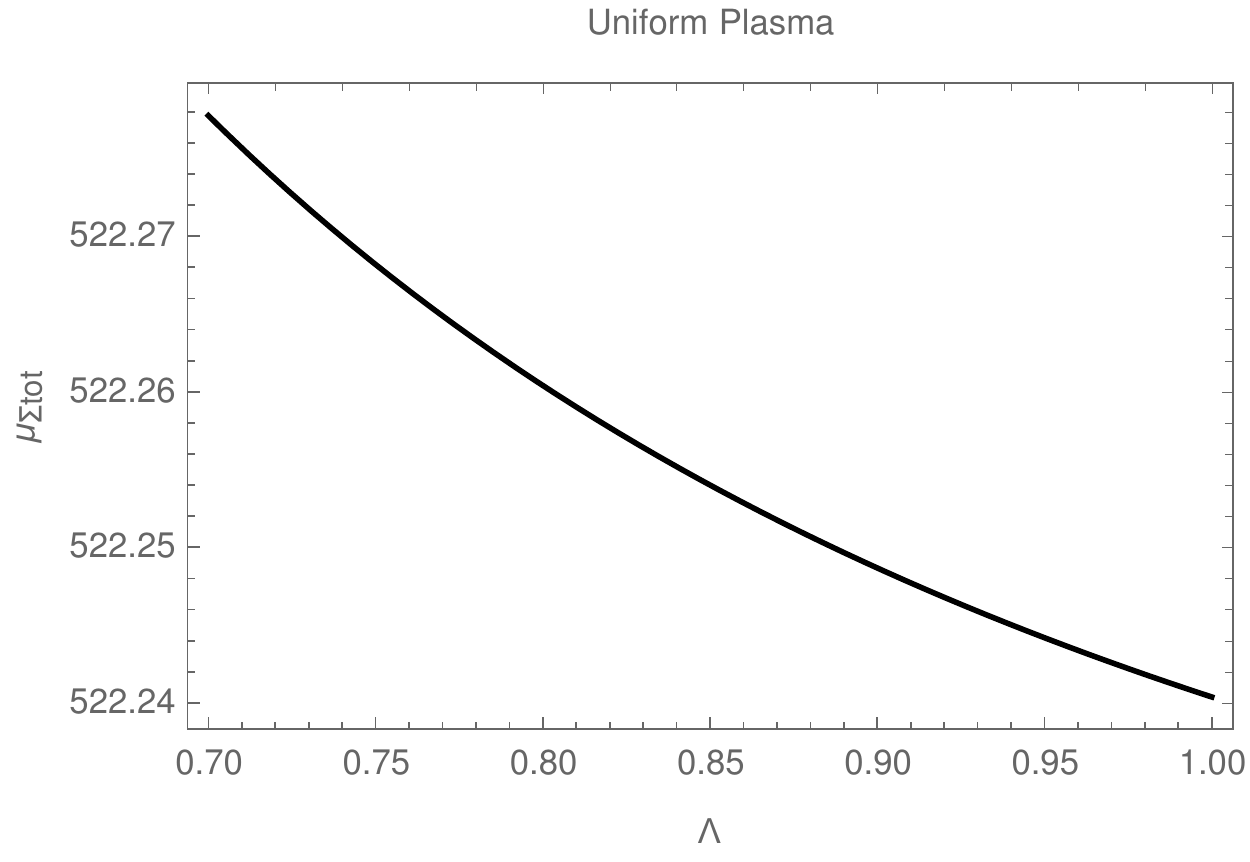}
d.\includegraphics[scale=0.65]{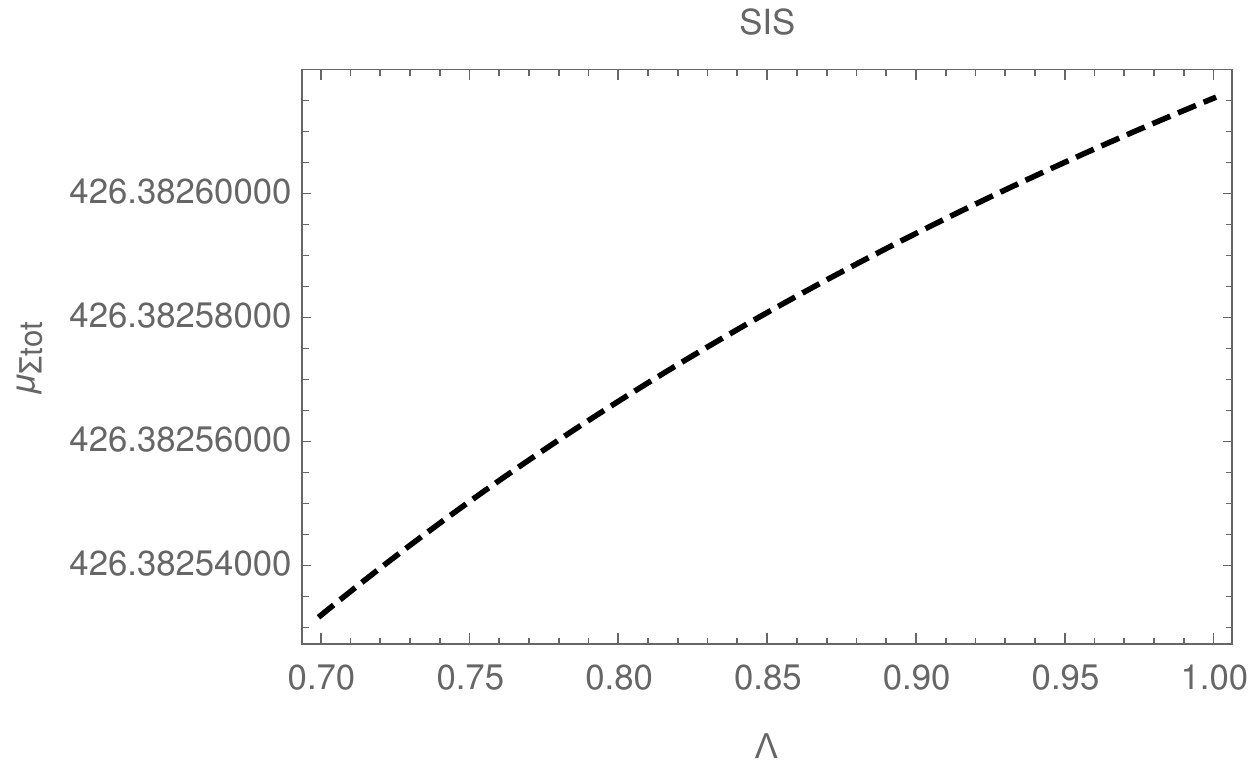}
\caption{(\textbf{a}) Plot of $\mu_{\Sigma tot}$ vs. $\Lambda$ when $\beta=0.001$ for uniform plasma. (\textbf{b}) Plot of $\mu_{\Sigma tot}$ vs. $\Lambda$ when $\beta=0.001$ for the \textbf{SIS} distribution. (\textbf{c}) Plot of $\mu_{\Sigma tot}$ vs. $\Lambda$ when $\beta=0.0001$ for unifomr plasma. (\textbf{d}) Plot of $\mu_{\Sigma tot}$ vs. $\Lambda$ when $\beta=0.0001$ for the \textbf{SIS} distribution. In all the figures we considered $\overline{D}_{ls}=10$, $\overline{D}_l=100$, $\overline{D}_s=110$, $\omega^2_e/\omega^2=\omega^2_c/\omega^2=0.5$, $\theta_E=0.001818$, and $\tilde{J}_r=0.3$. \label{fig17}}
\end{center}
\end{figure*}

In Figs. \ref{fig17}\textcolor{blue}{.b} and \ref{fig17}\textcolor{blue}{.d}, we plotted the behaviour of $\mu_{\Sigma tot}$ as a function of the boosted parameter $\Lambda$ for $\beta=0.001$ and $\beta=0.0001$ respectively. In contrast with the previous case (uniform plasma), we see that the total magnification increases as $\Lambda$ increases. On the other hand, note that for small values of $\beta$, the magnitude of $\mu_{\Sigma tot}$ increases: it changes, for example, from $42.6$ to $426.3$ when $\beta$ changes from $0.001$ to $0.0001$ respectively.    


\section{Conclusion}
\begin{figure*}[t]
\begin{center}
a.\includegraphics[scale=0.65]{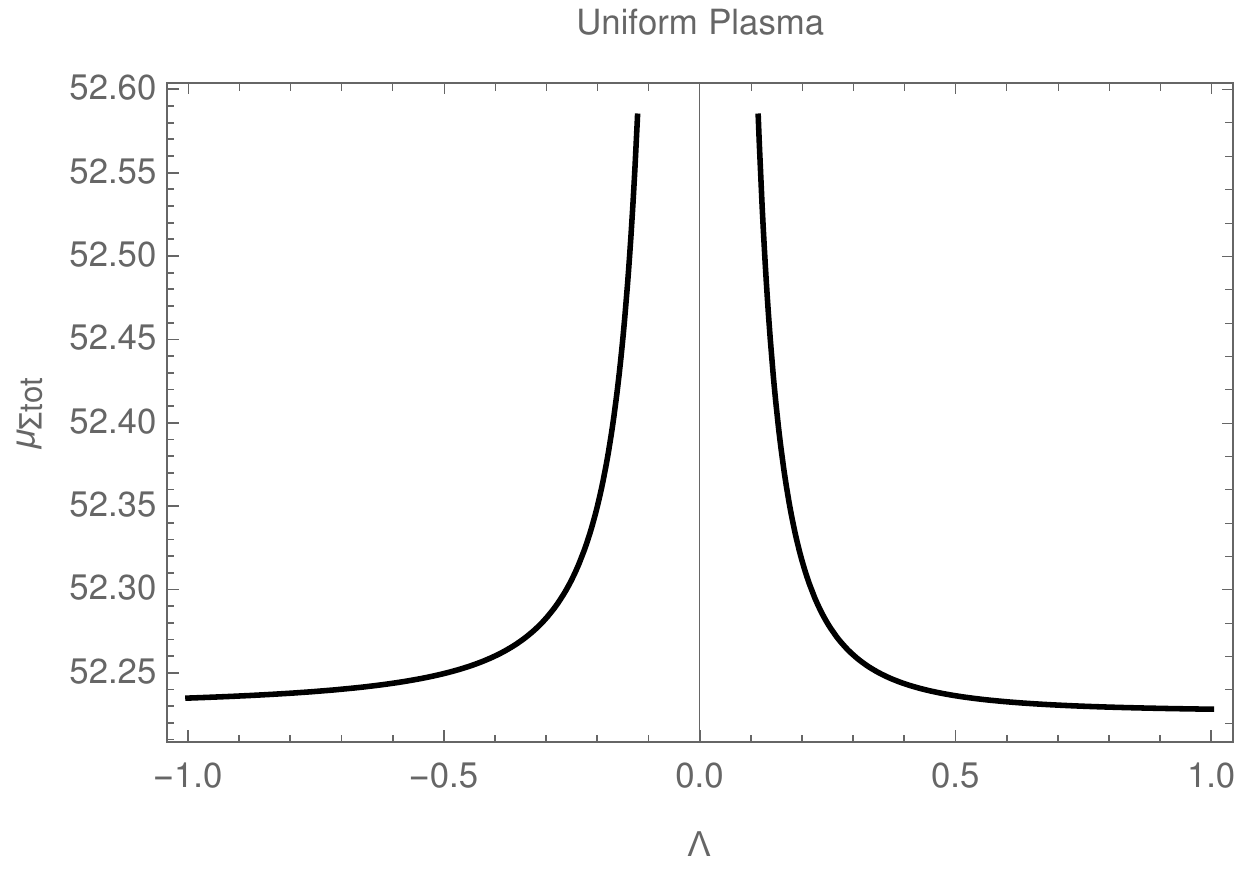}
b.\includegraphics[scale=0.65]{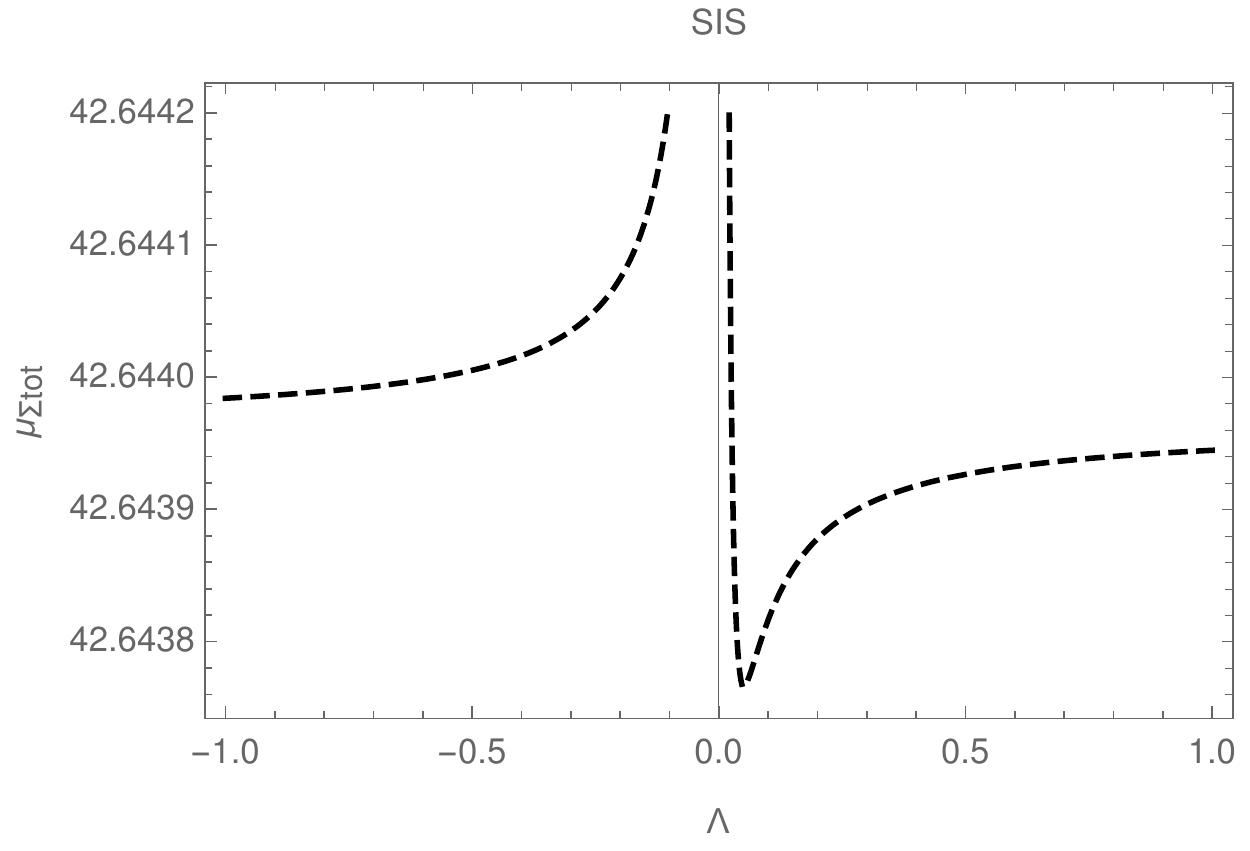}
\caption{(\textbf{a}) Plot of $\mu_{\Sigma tot}$ vs. $\Lambda$ when $\beta=0.001$ for uniform plasma. (\textbf{b}) Plot of $\mu_{\Sigma tot}$ vs. $\Lambda$ when $\beta=0.001$ for the \textbf{SIS} distribution. In all the figures we considered $\overline{D}_{ls}=10$, $\overline{D}_l=100$, $\overline{D}_s=110$, $\omega^2_e/\omega^2=\omega^2_c/\omega^2=0.5$, $\theta_E=0.001818$, and $\tilde{J}_r=0.3$. \label{fig18}}
\end{center}
\end{figure*}

In this work we have studied the deflection angle for the boosted Kerr metric in the presence of both homogeneous and non-homogeneous plasma, and in the latter case three different distributions have been  considered.\\

In Subsection~\ref{sec:nonrotating_case_uniformplasma} we investigated the behavior of the deflection angle for the non-rotating case in the presence of uniform plasma ($\omega_e= \text{costant}$) by considering small values of $v$. According to Eq.~(\ref{deflection_angle_nonrotating_case}) we found that $\hat{\alpha}_b$  does not dependent, at least at first order,  on the velocity $v$. It was also found that, after the approximation $1-n\ll \frac{\omega_e}{\omega}$, the deflection angle in Eq.~(\ref{delflection_angle_with_plasma_boosted_kerr_metric}) reduces to that obtained in~\cite{Kogan10} (see Eq.~(\ref{deflection_angle_nonrotating_case})). As a consequence, the optics for the non-rotating boosted Kerr metric is the same as Schwarzschild. In this sense, the bending of light, due to the presence of a uniform plasma, is greater than the Schwarzschild case in vacuum for values of $\omega^2_e/\omega^2$ smaller than unity.\\

In Subsection~\ref{sec:Deflection_angle_for_the_slowly_rotating_case}, we studied the rotating case by considering a uniform distribution. Following the ideas of \cite{Morozova13}, we found that the expression for the deflection angle $\hat{\alpha}_b$ in Eq.~(\ref{deflection_angle_rotating_case_homgeneous_plasma}) contains two terms: the Schwarzschild angle $\hat{\alpha}_{bS}$, and the contribution due to the dragging of the inertial frame ${\hat{\alpha}_{bD}}$. The result is quite similar to that of V.S Morozova et al.. However, in contrast with their result, Eq.~(\ref{deflection_angle_rotating_case_homgeneous_plasma}) also depends on  the parameter $\Lambda$. This dependence is shown in Fig.~\ref{fig5}. Form this figure we found that the smaller the values of $\Lambda$ (constrained to the interval $0<\Lambda\leq 1$) the greater is the deflection angle. In this sense, not only the dragging and the presence of a plasma, but also the motion of the black hole will contribute to the lensing. Therefore, since no effect was found in the previous case, we may concluded that $\hat{\alpha}_b$ depends on $v$ only when the dragging of the inertial frame takes place.\\

In Section~\ref{sec:Models}, we consider the deflection angle in terms of $\overline{b}$, $\Lambda$, and $\widetilde{J}_r$ for different distributions.  As shown in our figures, $\hat{\alpha}$ is affected by the presence of plasma and is greater when compared with vacuum and uniform distributions. Furthermore, we found again that $\hat{\alpha}$ increases not only due to the dragging, but also when small values of the boosted parameter $\Lambda$ are considered.\\

In this work, we also found some important constraints for two of the models. In the case of \textbf{NSIS}, for example, the radius of the core $r_c$ must have values greater than $6M$. If the core radius is smaller than this limit the deflection angle becomes negative at some point and will not agree with the usual behavior when $b\rightarrow\infty$. On the other hand, regarding the \textbf{PGC}, we found that $s$ must be different from $-1$ or $-3$ as can be seen from Eq.~(\ref{contributions_PGC}). Nevertheless, this condition is fulfilled since we consider positive values of $s<<1$. \\

No important difference between the models was found when the deflection angle was considered. In the case of \textbf{SIS} and \textbf{NSIS}, for example, the behavior was very similar. Therefore, under the weak field approximation, it is not possible to distinguish these two distributions. Nevertheless, the deflection angle is affected considerably when we consider a plasma in a galaxy cluster. The values of the deflection angle are greater than those obtained with the other two models. This behavior is clearly shown in Fig.~\ref{fig15}. Furthermore, according to Fig.~\ref{fig16}, we found that the deflection angle is affected by the plasma when the \textbf{PGC} distribution is considered.\\

Finally, in section \ref{sec:magnification}, as an application, we compute the total magnification for uniform and \textbf{SIS} plasma distributions. According to Fig. \ref{fig17} we conclude that, for small values of $v$ ($0.7\leq\Lambda\leq1$), the the total magnification is grater when the uniform plasma distribution is considered. For example, in the case of uniform distribution (considering $\beta=0.001$), we see that $\mu_{\Sigma tot}\approx52.22$. Nevertheless, for the \textbf{SIS} distribution, we found that $\mu_{\Sigma tot}\approx42.64$. A similar behaviour occurs when $\beta=0.0001$. Furthermore, it is important to point out that the total magnification has small changes in both distributions: $\mu_{\Sigma tot}$ ranges from $52.2285$ to $52.2305$ for the uniform plasma, and from $42.643938$ to $42.643944$ in the \textbf{SIS}. The change is very small for the last distribution.    \\

On the other hand, when we compare both models (uniform and \textbf{SIS} plasma distributions), we see that the behaviour of the total magnification is different (see figures \ref{fig18}\textcolor{blue}{.a} and \ref{fig18}\textcolor{blue}{.b}). In the case of the uniform plasma distribution, for example, when the boosted Kerr Black hole is moving towards ($\Lambda>0$) or away ($\Lambda<0$) from the observer the behaviour is very similar (there is a small difference when $\Lambda\rightarrow -1$ and $\Lambda\rightarrow 1$). However, when we consider the \textbf{SIS} distribution, the behaviour is not symmetric. This behaviour is due to cinematic effects. In this sense, when the magnification is considered, it would be possible to distinguish both models.


\begin{acknowledgements}

Authors thank the anonymous Referees for carefully reading the manuscript and for their value suggestions. We also want to thank Prof. N. Dadhich for value discussion. This work was supported by the National Natural Science Foundation of China (Grant No. U1531117) and Fudan University  (Grant No. IDH1512060). C.A.B.G. also acknowledges support from the China Scholarship Council (CSC), Grant No. 2017GXZ019022. C.B. also acknowledges the support from the Alexander von Humboldt Foundation. The research is supported in part by Grant No. VA-FA-F-2-008 and
No.YFA-Ftech-2018-8 of the Uzbekistan Ministry
for Innovation Development, by the Abdus Salam International
Centre for Theoretical Physics through Grant No. OEA-NT-01 and
by Erasmus+ exchange grant between Silesian
University in Opava and National University of Uzbekistan.

\end{acknowledgements}

\section*{Appendix I: Transformation to cartesian coordinates}
The transformation relations for the non rotating case ($a=0$) are (see \cite{Visser07})
\begin{equation}
\label{transformation_relation_nonrotating}
\begin{aligned}
\overline{t}&=t\\
\overline{x}&=r\sin\theta\cos\phi\\
\overline{y}&=r\sin\theta\sin\phi\\
\overline{z}&=r\cos\theta.\\
\end{aligned}
\end{equation}
Therefore, the Jacobian matrix has the form
\begin{equation}
\label{Jacobian_Matrix}
\mathbf{J}=\left(\begin{array}{cccc}
1&0&0&0\\
0&\cos\phi\sin\theta&r\cos\phi\cos\theta&-r\sin\phi\sin\theta\\
0&\sin\phi\sin\theta&r\sin\phi\cos\theta&r\cos\phi\sin\theta\\
0&\cos\theta&-r\sin\theta&0
\end{array}
\right).
\end{equation}
We are seeking for expressions of the form 
\begin{equation}
\label{transformation}
dx^\mu=\frac{\partial x^\mu}{\partial \overline{x}^\nu}d\overline{x}^\nu.
\end{equation}
In the last expression, $x^\mu$ denotes the Boyer-Lindquist coordinates ($t$, $r$, $\theta$, $\phi$) and $\overline{x}^\nu$ denotes the Cartesian coordinates ($\overline{t}$, $\overline{x}$, $\overline{y}$, $\overline{z}$). According to Eq.~(\ref{transformation}), the Jacobian for the inverse transformation has the form
\begin{equation}
\label{inverse_Jacobian}
\mathbf{J}^{-1}=\left(\frac{\partial x^\mu}{\partial \overline{x}^\nu}\right).
\end{equation} 
In order to find $\overline{J}$, we use the well known relation (see \cite{Levi-Civita61,Lovelock12}.) 
\begin{equation}
\mathbf{J}\times \mathbf{J}^{-1}=\mathbf{I}.
\end{equation}
Thus, the inverse transformation is 
\begin{equation}
\label{inverse_Jacobian_Matrix}
\mathbf{J}^{-1}=\left(\begin{array}{cccc}
1&0&0&0\\\\
0&\cos\phi\sin\theta&\sin\phi\sin\theta&\cos\theta\\\\
0&\frac{\cos\phi\cos\theta}{r}&\frac{\sin\phi\cos\theta}{r}&-\frac{\sin\theta}{r}\\\\
0&-\frac{\sin\phi}{r\sin\theta}&\frac{\cos\phi}{r\sin\theta}&0
\end{array}
\right),
\end{equation}
and, 
\begin{equation}
\label{inverse_transformation}
\begin{aligned}
dt&=d\overline{t}\\
dr&=\cos\phi\sin\theta d\overline{x}+\sin\phi\sin\theta d\overline{y}+\cos\theta d\overline{z}\\
d\theta&=\frac{\cos\phi\cos\theta}{r} d\overline{x}+\frac{\sin\phi\cos\theta}{r} d\overline{y}-\frac{\sin\theta}{r} d\overline{z}\\
d\phi&= -\frac{\sin\phi}{r\sin\theta} d\overline{x}+\frac{\cos\phi}{r\sin\theta} d\overline{y}.\\
\end{aligned}
\end{equation}
Then, after substitution in Eq.~(\ref{line_element_in_weak_limit_in_Boyer_lindquist_coordinates}) and taking into account that $dt=d\overline{t}$, the line element reduces to equation 
\begin{equation}
\begin{aligned}
\label{line_element_non_rotating_case_weak_fieldI}
ds^2&=ds^2_0+h_{11}d\overline{x}^2+h_{12}d\overline{x}d\overline{y}
+h_{13}d\overline{x}d\overline{z}\\
&+h_{22}d\overline{y}^2+h_{23}d\overline{y}d\overline{z}+\underbrace{\frac{2M}{r}}_{h_{00}}dt^2\\
&+d\overline{z}^2\underbrace{\left(\frac{2M}{r}\cos^2\theta-2v\cos\theta\sin^2\theta\right)}_{h_{33}},
\end{aligned}
\end{equation}
where  
\begin{equation}
\begin{aligned}
h_{11}&=-2v(\cos^2\phi\cos^3\theta+\sin^2\phi\cos\theta)\\
h_{12}&=4v(2\cos\phi\sin\phi\cos\theta-\cos\phi\sin\phi\cos^3\theta)\\
h_{13}&=4\left(\frac{M\cos\phi\cos\theta\sin\theta}{r}+v\cos\phi\cos^2\theta\sin\theta\right)\\
h_{22}&=2\left[\frac{M\sin^2\phi\sin^2\theta}{r}-v\left(\sin^2\phi\cos^3\theta+2\cos^2\phi\cos\theta\right)\right]\\
h_{23}&=4v\sin\phi\cos^2\theta\sin\theta\\
\end{aligned}
\end{equation}
For $v=0$, the line element in Eq.~(\ref{line_element_non_rotating_case_weak_fieldI}) reduces to the Schwarzschild case obtained in~\cite{Kogan10}.
 
\section*{Appendix II: Plasma distributions integrals}
\subsection*{Integrals in uniform plasma non-rotating case:}
The first integral in equation (\ref{deflection_angle_nonrotating_case}) is 
\begin{equation}
\int^\infty_{-\infty}\frac{dz}{(b^2+z^2)^\frac{3}{2}}=2\int^\infty_{0}\frac{dz}{(b^2+z^2)^\frac{3}{2}}=\frac{2}{b^2}
\end{equation}
\subsection*{Integrals in SIS:}
From Eqs.~(\ref{non_uniform_plasma}) and (\ref{plasma_frequency_SIS}) and the well known property of the $\Gamma$-function \cite{Gradshteyn07} 
\begin{equation}
\label{propertie_gamma_function}
\int^\infty_0\frac{dz}{(z^2+b^2)^{\frac{h}{2}+1}}=\frac{1}{hb^{h+1}}\frac{\sqrt{\pi}\Gamma\left(\frac{h}{2}+\frac{1}{2}\right)}{\Gamma\left(\frac{h}{2}\right)},
\end{equation}
the integrals of $\hat{\alpha}_{S2}$, $\hat{\alpha}_{S3}$, $\hat{\alpha}_{B2}$ and $\hat{\alpha}_{B3}$ are respectively 
\begin{equation}
\begin{aligned}
I_{S2}&=\int^{\infty}_0\frac{dz}{(b^2+z^2)^{\frac{3}{2}+1}}=\frac{\sqrt{\pi}}{3b^4}\frac{\Gamma(2)}{\Gamma(3/2)}=\frac{2}{3b^4}\\
I_{S3}&=\int^\infty_0\frac{dz}{(b^2+z^2)^2}=\frac{\sqrt{\pi}}{2b^3}\frac{\Gamma(3/2)}{\Gamma(1)}=\frac{\pi}{4b^3}\\
I_{B2}&=\int^{\infty}_0\frac{dz}{(b^2+z^2)^{\frac{3}{2}+1}}=\frac{2}{3b^4}\\
I_{B3}&=\int^{\infty}_0\frac{dz}{(b^2+z^2)^{\frac{5}{2}+1}}=\frac{\sqrt{\pi}}{5b^6}\frac{\Gamma(3)}{\Gamma(5/2)}=\frac{8}{15b^6}\\
\end{aligned}
\end{equation}
\subsection*{Integrals in NSIS:}
The integrals of $\hat{\alpha}_{S2}$, $\hat{\alpha}_{S3}$, $\hat{\alpha}_{B2}$ and $\hat{\alpha}_{B3}$, after substitution of Eq.~(\ref{plasma_frequency_SIS}) in (\ref{non_uniform_plasma}), are respectively 
\begin{equation}
\label{alpha_6_nonsingular_isothermal_sphere}
\begin{aligned}
\overline{I}_{S2}&=\int^\infty_0\frac{dz}{(z^2+b^2+r^2_c)(b^2+z^2)^\frac{3}{2}}\\
&=\frac{1}{b^2r^2_c}-\frac{{\rm arctanh}\left(\frac{r_c}{\sqrt{b^2+r^2_c}}\right)}{r^3_c\sqrt{b^2+r^2_c}}\\\\
\overline{I}_{S3}&=-\int^\infty_0\frac{dz}{(z^2+b^2+r^2_c)^2}\\
&=-\frac{\sqrt{\pi}}{2(b^2+r^2_c)^\frac{3}{2}}\frac{\Gamma(3/2)}{\Gamma(1)}\\\\
\overline{I}_{B2}&=\overline{I}_{S2}\\\\
\overline{I}_{B_3}&=\int^\infty_0\frac{dz}{(z^2+b^2+r^2_c)(b^2+z^2)^\frac{5}{2}}\\
&=\frac{2r^2_c-3b^2}{3r^4_cb^4}+\frac{{\rm arctanh}\left(\frac{r_c}{\sqrt{b^2+r^2_c}}\right)}{r^5_c\sqrt{r^2_c+b^2}}
\end{aligned}
\end{equation}

\subsection*{Integrals in PGC:}
The integrals of $\hat{\alpha}_{S2}$, $\hat{\alpha}_{S3}$, $\hat{\alpha}_{B2}$ and $\hat{\alpha}_{B3}$, after substitution of Eq.~(\ref{plasma_frequency_SIS}) in (\ref{non_uniform_plasma}), are respectively 
\begin{equation}
\label{alpha_6_nonsingular_isothermal_sphere}
\begin{aligned}
\widetilde{I}_{S2}&=\int^\infty_0\frac{dz}{(b^2+z^2)^{\frac{s+1}{2}+1}}=\frac{\sqrt{\pi}}{(s+1)b^{s+2}}\frac{\Gamma(\frac{s}{2}+1)}{\Gamma(\frac{s+1}{2})}\\\\
\widetilde{I}_{S3}&=-\int^\infty_0\frac{dz}{(b^2+z^2)^{\frac{s}{2}+1}}=\frac{\sqrt{\pi}}{sb^{s+1}}\frac{\Gamma(\frac{s+1}{2})}{\Gamma(\frac{s}{2})}\\\\
\widetilde{I}_{B2}&=\widetilde{I}_{S2}\\\\
\widetilde{I}_{B_3}&=\int^\infty_0\frac{dz}{(b^2+z^2)^{\frac{s+3}{2}+1}}=\frac{\sqrt{\pi}}{(s+3)b^{s+4}}\frac{\Gamma(\frac{s+4}{2})}{\Gamma(\frac{s+3}{2})}\\
\end{aligned}
\end{equation}

\bibliographystyle{spphys}       
\bibliography{gravreferences}   

\end{document}